\documentclass[fleqn,usenatbib]{mnras}

\usepackage[T1]{fontenc}
\usepackage{ae,aecompl}

\usepackage{graphicx}	
\usepackage{amsmath}	
\usepackage{amssymb}	
\usepackage[T1]{fontenc} 
\usepackage{academicons}
\usepackage{xcolor}
\usepackage{hyperref}

\newcommand{\ha}{H$\alpha$}
\newcommand{\hb}{H$\beta$}

\newcommand{\mic}{$\mu$m}
\newcommand{\orcid}[1]{\href{https://orcid.org/#1}{\textcolor[HTML]{A6CE39}{\aiOrcid}}}

\def\ltsima{$\buildrel<\over\sim$}
\def\lsim{\lower.5ex\hbox{\ltsima}~}
\def\gtsima{$\buildrel>\over\sim$}
\def\gsim{\lower.5ex\hbox{\gtsima}~}

\def\luv{$L_{\rm UV}$}

\def\sfrha{SFR$_{H\alpha}$}
\def\sfruv{SFR$_{UV}$}
\def\sfrir{SFR$_{IR}$}

\def\msolyr{M$_{\odot}$~yr$^{-1}$}
\def\msol{M$_{\odot}$}
\def\mstar{M$_{\star}$}

\def\teff{\ifmmode T_{\rm eff} \else $T_{\mathrm{eff}}$\fi}

\def\lya{Ly$\alpha$} 
\def\ha{H$\alpha$} 
\def\hb{H$\beta$}

\def\ebvs{$E_{B-V,\mathrm{stars}}$}
\def\ebvg{$E_{B-V,\mathrm{gas}}$}

\def\fesc{$f_{esc}$}

\def\ergscm{erg~s$^{-1}$~cm$^{-2}$}
\def\cm2{cm$^{-2}$}

\def\ewo3{$EW_{\mathrm{[O\textsc{iii}]}}$}
\def\ewha{$EW_{\mathrm{H}\alpha}$}

\def\oiii{O{\sc iii}}

\def\nh{\ifmmode N_{\mathrm{HI}}\else $N_{\mathrm{HI}}$\fi}

\def\vexp{\ifmmode v_{\rm exp} \else v$_{\rm exp}$\fi}
\def\taua{\ifmmode \tau_{a}\else $\tau_{a}$\fi}

\def\xiion{$\xi_{\mathrm{ion}}$}
\def\nion{$\dot{N}_{\mathrm{ion}}$}

\usepackage{newtxtext,newtxmath}

\title[Bursty star formation in low-mass galaxies]{The star formation burstiness and ionizing efficiency of low-mass galaxies}

\author[Atek et al.]{Hakim Atek$^{1}$\thanks{E-mail: hakim.atek@iap.fr},
Lukas J. Furtak$^{1}$,
Pascal Oesch$^{2,3}$,
Pieter van Dokkum$^{4}$,
Naveen Reddy$^{5}$,
\newauthor Thierry Contini$^{6}$,
Garth Illingworth$^{7}$,
Stephen Wilkins$^{8}$
\\
$^{1}$Institut d'astrophysique de Paris, CNRS, Sorbonne Universit\'e, 98bis Boulevrad Arago, 75014, Paris, France\\
$^{2}$Department of Astronomy, University of Geneva, 51 Ch. Pegasi, 1290 Versoix, Switzerland\\
$^{3}$Cosmic Dawn Center (DAWN), Niels Bohr Institute, University of Copenhagen, Jagtvej 128, K\o benhavn N, DK-2200, Denmark \\
$^{4}$Yale Center for Astronomy and Astrophysics, Yale University, New Haven, CT 06511, USA\\
$^{5}$Department of Physics \& Astronomy, University of California, Riverside, CA 92521, USA \\
$^{6}$ Institut de Recherche en Astrophysique et Plan\'etologie, CNRS,  Universit\'e de Toulouse, 14, avenue Edouard Belin, F-31400 Toulouse, France \\
$^{7}$ UCO/Lick Observatory, University of California, Santa Cruz, CA 95064, USA\\
$^{8}$Astronomy Centre, Department of Physics and Astronomy, University of Sussex, Brighton, BN1 9QH, UK\\
}

\date{Accepted XXX. Received YYY; in original form ZZZ}

\pubyear{2022}

\begin{document}
\label{firstpage}
\pagerange{\pageref{firstpage}--\pageref{lastpage}}
\maketitle

\begin{abstract}
We investigate the burstiness of star formation and the ionizing efficiency of a large sample of galaxies at $0.7 < z < 1.5$ using {\em HST} grism spectroscopy and deep ultraviolet (UV) imaging in the GOODS-N and GOODS-S fields. The star formation history (SFH) in these strong emission line low-mass galaxies indicates an elevated star formation rate (SFR) based on the \ha\ emission line at a given stellar mass when compared to the standard main sequence. Moreover, when comparing the \ha\ and UV SFR indicators, we find that an excess in \sfrha\ compared to \sfruv\ is preferentially observed in lower-mass galaxies below $10^{9}$ \msol, which are also the highest-EW galaxies. These findings suggest that the burstiness parameters of these strong emission line galaxies may differ from those inferred from hydrodynamical simulations and previous observations. For instance, a larger burstiness duty cycle would explain the observed \sfrha\ excess. We also estimate the ionizing photon production efficiency \xiion, finding a median value of Log(\xiion/erg$^{-1}$ Hz)$=24.80 \pm 0.26$ when adopting a Galactic dust correction for \ha\ and an SMC one for the stellar component. We observe an increase of \xiion\ with redshift, further confirming similar results at higher redshifts. We also find that \xiion\ is strongly correlated with \ewha, which provides an approach for deriving \xiion\ in early galaxies. We observe that lower-mass, lower-luminosity galaxies have a higher \xiion. Overall, these results provide further support for faint galaxies playing a major role in the reionization of the Universe.
\end{abstract}




\section{Introduction}

There is a growing interest in the study of low-mass galaxies at all epochs, both on the theoretical and observational fronts. Until recently, most of the efforts were dedicated to characterizing massive galaxy populations out to the highest redshifts, and little was known about high$-z$ star-forming galaxies with stellar masses below $10^{10}$ \msol. Near-infrared spectroscopic surveys from space, such as 3D-HST \citep{brammer12} and WISPS \citep{atek10}, and subsequent follow-ups with ground-based multi-object spectrographs \citep[e.g.,][]{wisnioki15,kriek15}, have given access to the main rest-frame optical emission lines, which enable us to derive the physical properties of the ISM in high-redshift galaxies. In particular, various studies have recently focused on investigating the star formation histories of low-mass galaxies and how they compare to their massive counterparts \citep[e.g.,][]{guo16,sparre17,velasquez20}.

In the classic picture of galaxy evolution in which the SFR varies smoothly with time, star-forming galaxies follow a well known relation between the SFR and the stellar mass \mstar\ \citep[e.g.,][]{brinchmann04,elbaz07,wuyts11,whitaker14}. It is believed that such a relation is  shaped by the complex interplay between several physical mechanisms that operate over long timescales (Giga years). For instance, the current SFR depends on the gas accretion and the merger history of the galaxy and star-formation suppression by the AGN feedback. While this relation is expected to hold for relatively massive galaxies (> $10^{10}$ \msol), it is unlikely that star formation quenching from AGN remains very efficient in low-mass galaxies. Supernova explosions become the most significant source of feedback processes. At a given redshift, low-mass galaxies appear to have a higher specific star formation rate (sSFR, which is the SFR per unit of stellar mass) than "normal" galaxies, and can double their total stellar mass on a very short timescale of 100 Myr \citep{rodighiero11, atek14c, boogaard18}. Star formation is expected to occur mostly through a stochastic process in these galaxies as a result of two competing processes: gas accretion feeding intense bursts of star formation and feedback from supernova explosions. Such variations are expected to occur on short time scales ($\sim 10$ Myr) and imply that galaxy observables might not be representative of the averaged properties of galaxies \citep{dominguez15}. Bursty star formation also has important implications on galaxy formation and evolution, as it may be responsible for altering dark matter density profiles \citep{chan15,read16,pelliccia20}.  

The burstiness of star formation can be investigated by comparing SFR(\ha) and SFR(UV), two indicators that trace different timescales: while the \ha\ nebular emission traces the ionizing radiation from short-lived massive O stars over few to ten Myr, the UV continuum traces a population of  longer-lived O, B and A stars over $\sim 100$ Myr \citep[see][for a review]{kennicutt12}. In the case of a constant star formation, the ratio between these two indicators will be close to unity after 100 Myr, whereas deviations are to be expected for rapidly varying star formation. Hydrodynamical simulations have shown that burstiness increases towards lower-mass galaxies \citep[e.g.][]{shen14,velasquez20}, which leads to a decreasing \sfrha/\sfruv\ ratio with decreasing stellar mass.

This is due to the relatively rapid response of SFR(\ha) to the change in instantaneous SFR compared to SFR(UV) in the same burst period. This is why the ratio between these two SFRs is an indicator of burstiness. It means that, within a 10 Myr period of an instantaneous SF burst, SFR(\ha) will increase rapidly, following closely the "true" SFR, whereas SFR(UV) will slowly increase, falling behind the "true" SFR. In this phase we will have \sfrha/\sfruv\ above unity \citep[see for example][]{sparre17}. After the burst, The FUV emission will decrease slowly, whereas the \ha\ emission fades quickly. Therefore, the \sfrha/\sfruv\ goes below unity after the burst episode. Similar results were also observed in local and high-redshift galaxies \citep{weisz12,guo16, emami19}, whereas no sign of burstiness was found by \citet{smit16}. \citet{faisst19} also find that more than half of their sample of $z \sim 4.5$ galaxies have an \sfrha/\sfruv\ ratio above unity.

Low-mass galaxies are also suspected to be major contributors to cosmic reionization. According to the very steep slope ($\alpha \sim -2$) of the faint-end of the UV luminosity function at $z > 6$ \citep[e.g.][]{atek15b, bouwens15, livermore17, ishigaki18, atek18}, the UV radiation budget is dominated by galaxies fainter than $M_{UV} = -17$ mag. One needs to convert this non-ionizing UV radiation to the ionizing emission below 912 \AA, which can be achieved by estimating the ionizing efficiency \xiion. It is a measure of the ionizing production rate relative to the UV luminosity density at 1500 \AA. The primary method to compute \xiion\ is to use a nebular recombination line such \ha\ to infer the ionizing photon production rate, which is then compared to UV luminosity. Therefore, this \xiion\ estimate is based on the same  \ha/UV ratio used to investigate the star formation burstiness. Indirect measurements of \ha\ have been obtained through the photometric excess in Spitzer IRAC bands to infer \xiion\ in $z \sim 4$ galaxies \citep[e.g.][]{smit16, lam19}. Spectroscopic measurements of \ha\ or \hb\ lines have also been conducted to infer \xiion\ of $z \sim 2-3$ galaxies. However, these studies either focused on relatively massive galaxies \citep[\mstar $=10^{9}-10^{11}$ \msol;][]{shivaei18}, a relatively small sample of galaxies \citep{emami20}, or SED fitting to infer the UV properties in bright \lya\ emitters \citep{nakajima16}. At higher redshift, \citet{stark15} used stellar population and photoionization models to fit rest-frame UV spectra of $z \sim 7$ galaxies and infer \xiion\ from the best-fit SED model.

In this work, we use a large sample of galaxies at $0.7 < z < 1.5$ reaching down to a stellar mass of $\sim 10^{8}$ \msol, with both spectroscopic measurements of \ha\ emission and deep UV imaging to o investigate the star formation burstiness. We used rest-frame optical spectra from 3DHST grism survey in the GOODS-S and GOODS-N fields and rest-frame UV imaging around 1500 \AA\ from HDUV program. Nebular attenuation was inferred from a subsample of galaxies with Balmer decrement measurements and the mass-attenuation relation. The continuum attenuation was computed from the UV continuum slope $\beta$. We also used these quantities to directly compute the ionizing efficiency.

The paper is structured as follows. In Section \ref{sec:obs}, we present the observational data used in the paper. In Section \ref{sec:sample}, we describe how the galaxy sample was selected and how dust attenuation was derived. The star formation burstiness is discussed in Section \ref{sec:burstiness}. We investigate its impact on the location of galaxies in the SFR-\mstar\ plane and how the ratio of SFR indicators \sfrha/\sfruv\ evolves towards low-mass galaxies. The method for deriving the ionizing efficiency is described in Section \ref{sec:xi}. We also discuss the impact of dust attenuation on the final \xiion\ results. Finally, we compare our estimate with a compilation of literature results and derive the redshift-evolution of \xiion. In Section \ref{sec:prop}, we explore the evolution of \xiion\ with the \ha\ end [\oiii] equivalent widths and the galaxies properties, such as the stellar mass and the UV magnitude. In Section \ref{sec:reionization}, we discuss the implications of star formation burstiness and \xiion\ estimates on the contribution of galaxies to cosmic reionization, before presenting our summary and conclusions in Section \ref{sec:conclusions}. Throughout the paper, magnitudes are in the AB system \citep{oke83} and we adopt a cosmology with H${_0} =70$ km s$^{-1}$ Mpc$^{-1}$, $\Omega_{\Lambda}=0.7$, and $\Omega_m=0.3$.

\section{Observations}
\label{sec:obs}
The spectroscopic data used in this paper were obtained from the data release v4.1.5 of the 3D-HST survey \citep{brammer12,momcheva16}. The program uses about 250 orbits of the {\em Hubble Space Telescope} ({\em HST}) to observe a total of 124 pointings in four of the CANDELS fields \citep{grogin11}. The release also include similar observations in the fifth field (GOODS-N) obtained by the AGHAST program (GO-11600). Observations consist of slitless spectra obtained with the G141 grism on the Wide Field Camera 3 (WFC3) and short direct imaging observations with the F140W filter. The near-infrared (NIR) spectra cover the wavelength range 1.1 to 1.65 $\mu$m at an instrumental resolution of R=130. For more details about the data reduction, we refer the reader to \citet{momcheva16}. We also use matched photometric catalogs constructed on the CANDELS and 3D-HST observations, which are complemented with ancillary data between 0.3 and 8 $\mu$m \citep[cf.][for details]{skelton14}. 

The matched emission line catalogs include multiple quality flags down to an AB magnitude of $JH=24$. Beyond this magnitude limit, we visually inspected the spectral fits of all objects in the catalog. Most of the discarded objects were due to line misidentification and prominent noise features.

The ultraviolet imaging data used here are part of the HDUV legacy survey (GO-13872), which targets 13 pointings in the CANDELS areas of the GOODS-North and GOODS-South fields. The program obtains deep observations of typically 4 orbits in F275W and 8 orbits in F336W filters of the UVIS channel of WFC3, reaching down to a $5\sigma$ depth of 27.5-28 mag. The data also include very deep observations over the Hubble Ultra Deep Field (HUDF) obtained as part of the UVUDF program \citep{teplitz13}. The data reduction procedure is detailed in the survey paper by \citet{oesch18}.

\section{The galaxy Sample}
\label{sec:sample}
\begin{figure*}
         \includegraphics[width=\columnwidth]{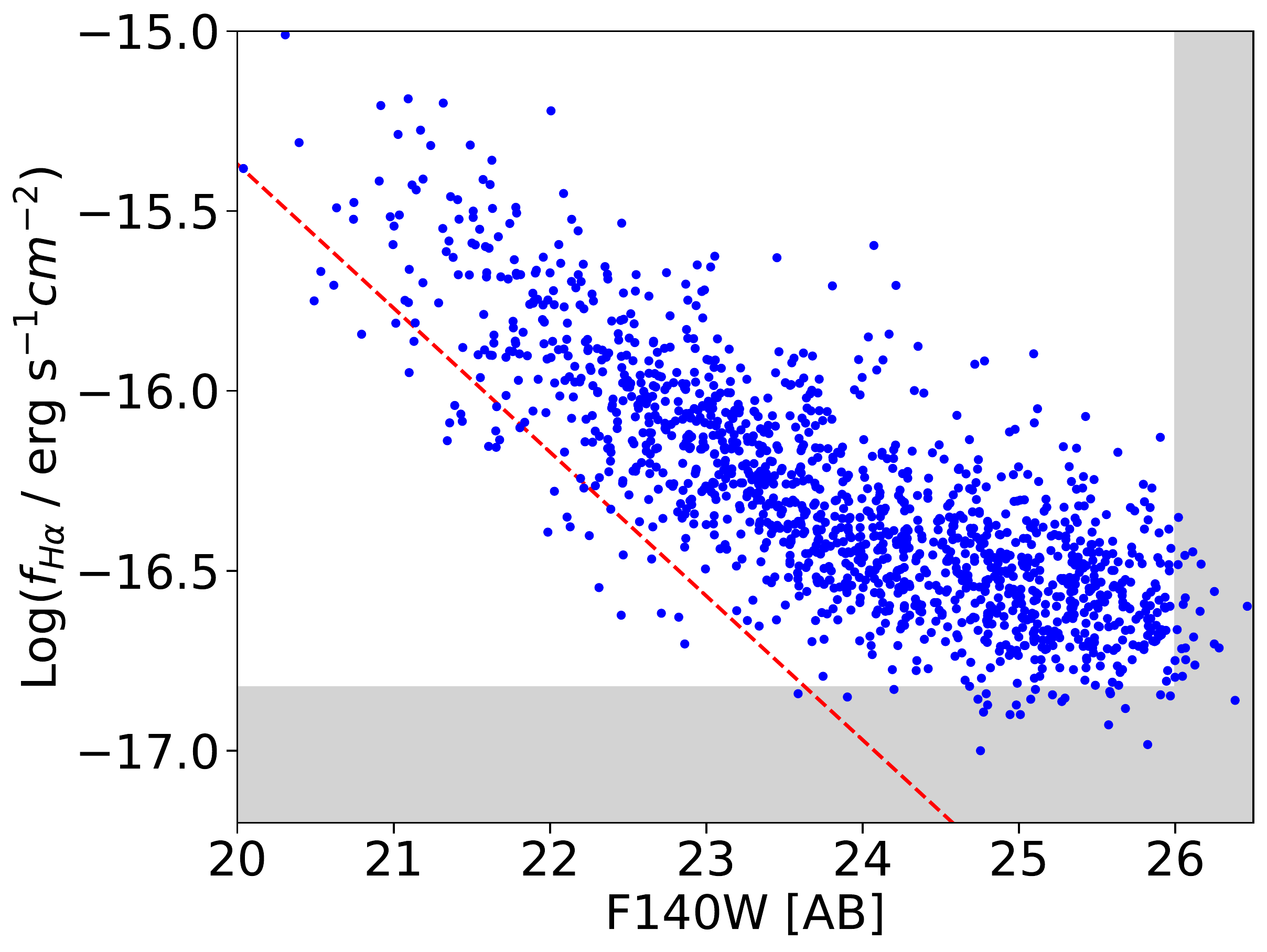}
         \hspace{0.5cm}
		\includegraphics[width=7cm]{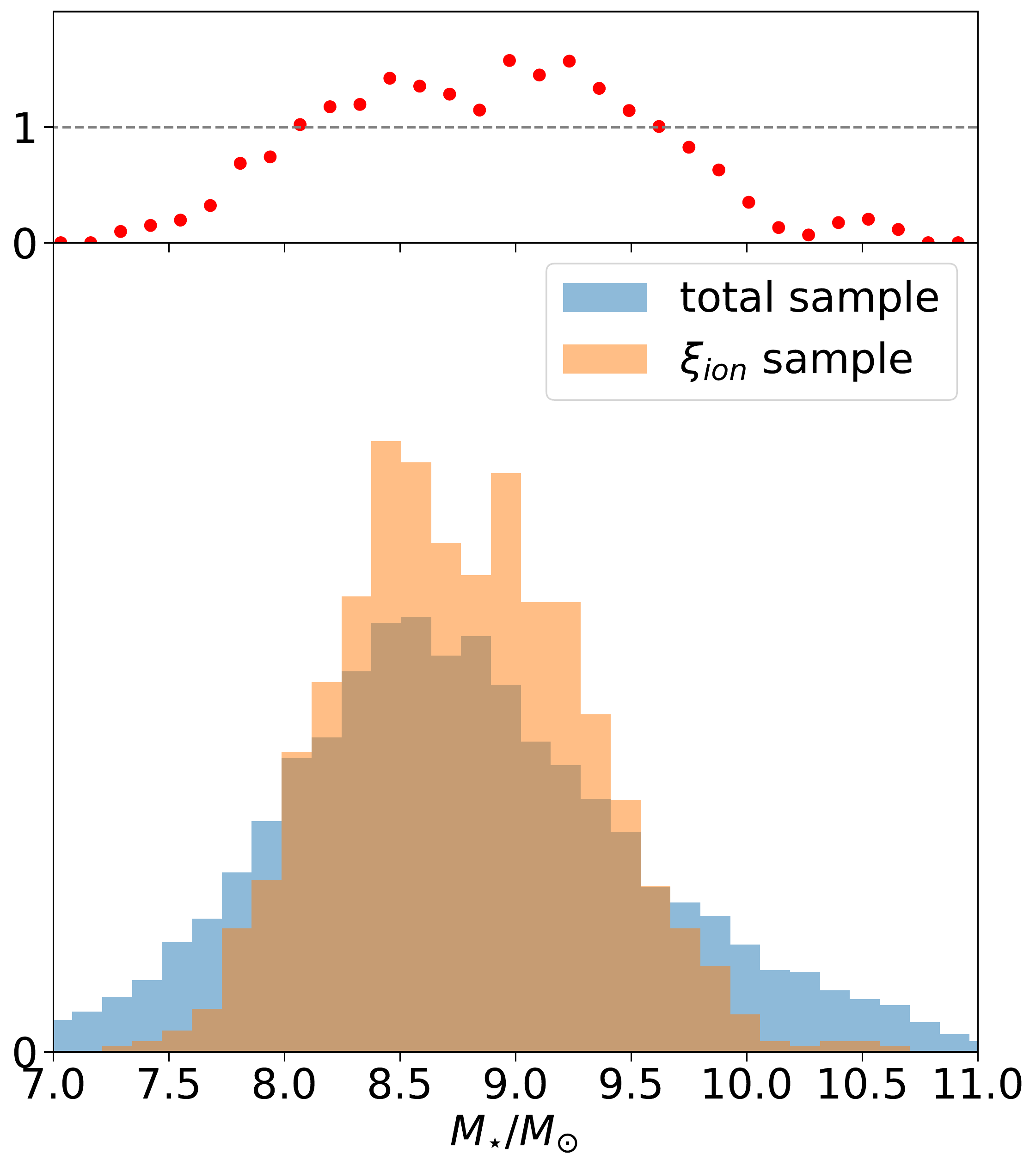}
    \caption{Completeness estimate of the sample selection.  {\bf Left:} \ha\ flux as a function of magnitude in the NIR F140W band for the galaxy sample selected for this study. The red dashed line represents \ewha = 80 \AA, the observed equivalent width cut used in our selection. The gray-shaded areas denote the \ha\ flux limit of $1.5 \times 10^{-17}$ \ergscm\ and the magnitude limit of F140W images at $AB=26$ mag. {\bf Right:} stellar mass distribution for the entire spectroscopic sample (in blue) and the selected sample (in orange). Both distributions are re-normalized. The top panel shows the ratio of the relative fractions of both samples. It shows a significant decrease in the completeness at the low (\mstar $\sim 10^{8}$ \msol) and high (\mstar $\sim 10^{9.5}$ \msol) stellar mass. Stellar masses are part of the 3D-HST data release and were originally estimated by \citet{skelton14}.}
    \label{fig:comp}
\end{figure*}

The combination of the grism spectroscopy and the UV imaging provides us with the opportunity to explore the star formation activity on different timescales and measure the ionizing efficiency of star forming galaxies at $z \sim 1$ as a function of their physical parameters. We have selected our sample based on high-significance \ha\ emission lines (above SNR=3) and equivalent widths higher than 80 \AA\ in the spectroscopic data, and SNR$>3$ continuum detection in the F336W filter of the UV data. The final sample contains a total of 1167 objects in the redshift range $0.7<z<1.5$. 

The NIR grism spectroscopy is particularly sensitive to strong nebular emission lines. Most of the objects in this sample will have relatively high equivalent widths, and in some cases, their continuum will remain undetected. Therefore, equivalent width measurements will be highly uncertain. We have re-computed the equivalent width of all the emission lines using the line flux and the broadband magnitude in the $JH_{140}$ images, correcting for the contribution of line flux to the observed magnitudes:
\begin{equation}
    EW_{cor} = \frac{F_{line}} {f_{\lambda, 140}  - \frac{\int f_{\lambda, line} d\lambda}{\int d\lambda} }
\end{equation}
where $f_{line}$ is the line flux  and $f_{\lambda, 140}$ is the flux density in the F140W filter.
While continuum-selected samples are limited by the depth of the imaging data, our emission-line selection favors high equivalent width galaxies showing enough contrast between the line and the continuum. The left panel of Figure \ref{fig:comp} shows the impact of this selection on the completeness in \ha\ flux as a function of magnitude. While the faint-end limits of both continuum and line fluxes are represented by the shaded area, the line flux limit increases towards brighter magnitudes. Beyond a magnitude $JH \sim 24$, the sample is defined not only by the equivalent width, but also by the \ha\ flux limit. For these faint galaxies, we will only select bright \ha\ emitters relative to their continuum magnitude. This impact of this selection bias will be in Section \ref{sec:burstiness}. Comparing the stellar mass distribution for the \ha\ sample and the parent sample, we can see on the right panel of Figure \ref{fig:comp} that the selected sample becomes incomplete towards low stellar masses, typically around Log(\mstar/\msol $= 8$), and at high masses around Log(\mstar/\msol $= 9.5$). This is due to our sample selection, which is based on SNR cuts on both \ha\ and UV emission, on one hand, and EW(\ha), on the other hand.    

\subsection{Dust attenuation correction} 
\label{sec:dust}

\begin{figure}
        \includegraphics[width=\columnwidth]{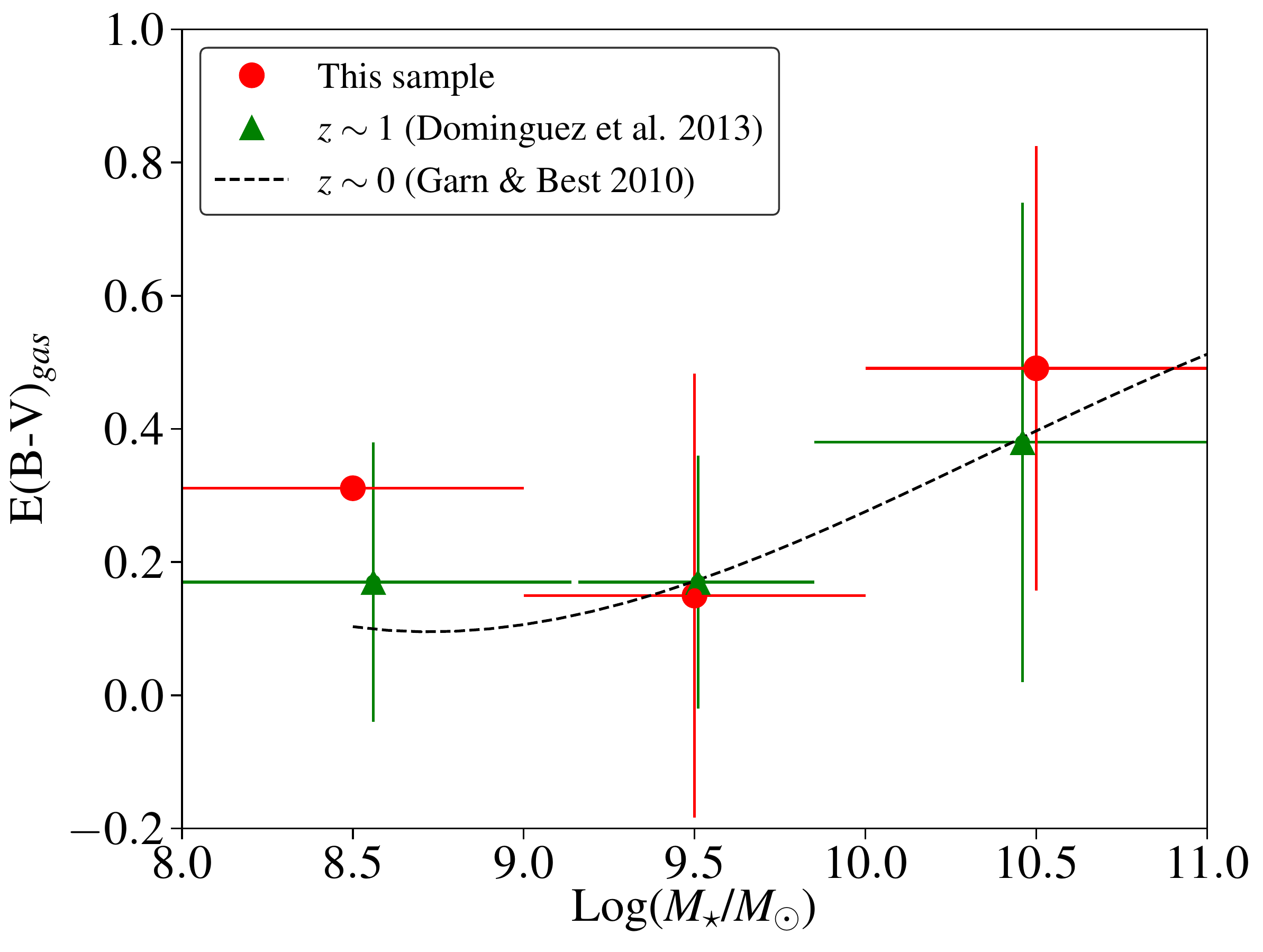}
    \caption{Nebular attenuation as a function of the stellar mass. The reddening E(B-V)$_{gas}$ is computed using the Balmer decrement of a subsample of galaxies binned in stellar mass. Our data points (red circles) are plotted together with the results of \citet{dominguez13} (green triangles). The dashed line represents the relation derived for local galaxies by \citet{garn10}.}
    \label{fig:ebv_mass}
\end{figure}

To correct the parameter measurements for dust attenuation, we consider the \ha\ and UV continuum emission independently. First, we compute the nebular attenuation from the Balmer decrement \ha/\hb\ for a subsample of 10 galaxies over a restricted redshift range of $1.3 < z < 1.51$, where both \ha\ and \hb\ lines are visible in the G141 grism, and we examine the trends as a function of stellar mass and \ha\ luminosity. Figure \ref{fig:ebv_mass} shows a good agreement with the results of \citet{dominguez13} who used a stack of $0.75<z<1.5$ emission-line spectra to derive Balmer decrements as a function of stellar mass. It also follows the trend derived for local galaxies \citep{brinchmann04,garn10} with no sign of evolution with redshift, which was also observed at $z > 1$ \citep[e.g.][]{sobral12,dominguez13}. We apply the relation derived by \citet{dominguez13} to the entire galaxy sample to estimate the nebular attenuation. 

As for the UV luminosity, we compute the UV slope over a rest-frame spectral range of $1300-3300$ \AA\ by performing a multi-band fitting to the relation $f_{\lambda} \propto \lambda^{\beta}$. Then we derive the attenuation by adopting an intrinsic UV slope of $\beta_{0} = -2.62$ \citep{reddy18}. The intrinsic UV slope is sensitive to the star formation history (SFH) of galaxies, which in turns affects the computed stellar attenuation. Galaxies with higher specific star formation rate have bluer UV colors due to a lower contribution of older stellar populations. The present sample has relatively high sSFRs and strong emission lines, with median values of sSFR=$10^{-8.4}$ yr$^{-1}$ and EW$_{\rm rest}$(\ha)=160 \AA\ (cf. Fig. \ref{fig:histo}). We show in Appendix \ref{app:a3} the effects of adopting the standard value of $\beta_{0}=-2.23$ of \citet{meurer99}.  
\begin{figure}
    \centering
    \includegraphics[width=\columnwidth]{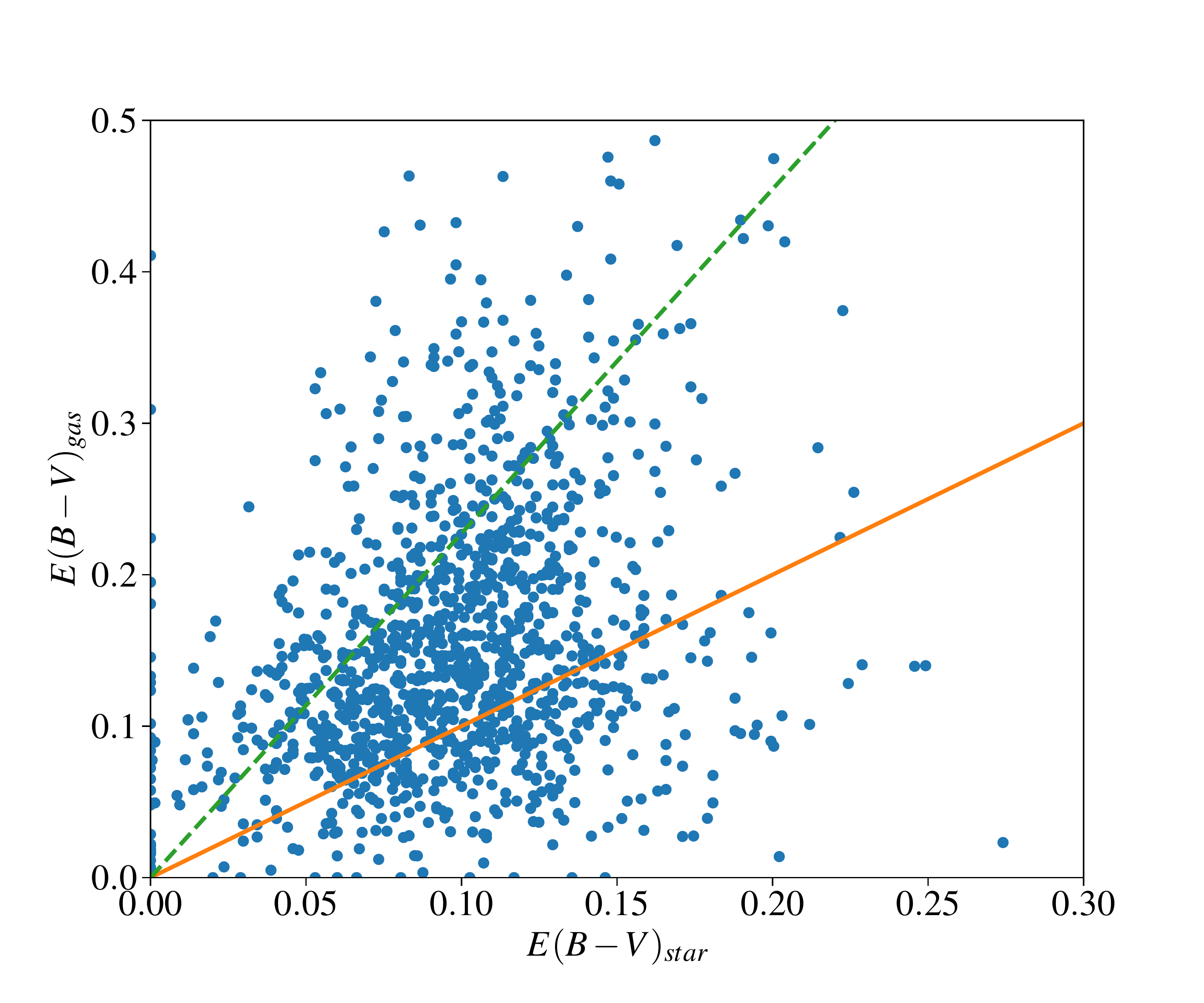}
    \caption{ The nebular dust reddening {\em vs} the stellar dust reddening. The reddening values are computed using the main combination of dust laws: the \citet{cardelli89} attenuation law for \ebvg\ and the SMC extinction law for \ebvs. The solid orange line denotes the 1:1 relation, while the dashed blue line marks the \citet{calzetti00} relation \ebvs = 0.44\ebvg.}
    \label{fig:ebvg_ebvs}
\end{figure}

\begin{table}
    \centering
    \begin{tabular}{l|c|c|c}
    \hline
    Emission & Curve  & E(B-V) & Correction Factor  \\
    \hline
    \ha\ & Cardelli     &   0.14      &       1.40  \\
    \ha\ & Calzetti     &   0.12      &      1.50 \\
    UV  & SMC           &    0.10     &      3.3    \\ 
     UV & Calzetti     &    0.23     &      11.3  \\
     \hline
    \end{tabular}
    \caption{The importance of dust correction for the \ha\ and UV emissions. We list the reddening estimates and the correction factors for the different extinction/attenuation laws considered in this work.The E(B-V) values represent the nebular and the stellar reddening for the \ha\ and UV fluxes, respectively}.
    \label{tab:dust}
\end{table}

Many efforts have been undertaken to investigate the attenuation law in high-redshift galaxies. There has been some evidence for a different calibration compared to local galaxies, possibly due to their younger age, lower metallicities or their different emission line properties in general \citep{debarros14,price14,reddy15,smit16,cowie16}. In Section \ref{sec:burstiness} we adopt the \citet{cardelli89} curve for \ha\ emission and the SMC extinction curve \citep{gordon03} for the UV continuum. When computing absolute values such as the ionizing efficiency (Section \ref{sec:xi}) we explore the effects of adopting two laws for the nebular attenuation, namely the \citet{calzetti00} and \citet{cardelli89} curves. Regarding the stellar continuum, we also compare the SMC extinction curve \citep{gordon03} with the \citet{calzetti00} attenuation curve in computing E(B-V)$_{s}$, keeping the same intrinsic slope $\beta_{0} = -2.62$. Table \ref{tab:dust} shows the resulting correcting factors for \ha\ and UV luminosities when adopting different dust laws. We also show in Figure \ref{fig:ebvg_ebvs} the relation between \ebvg\ and \ebvs. An important dispersion is observed, with most of the sample lying between the 1:1 ratio and the 0.44 factor of \citet{calzetti00}.

\begin{figure}
        \includegraphics[width=0.5\columnwidth]{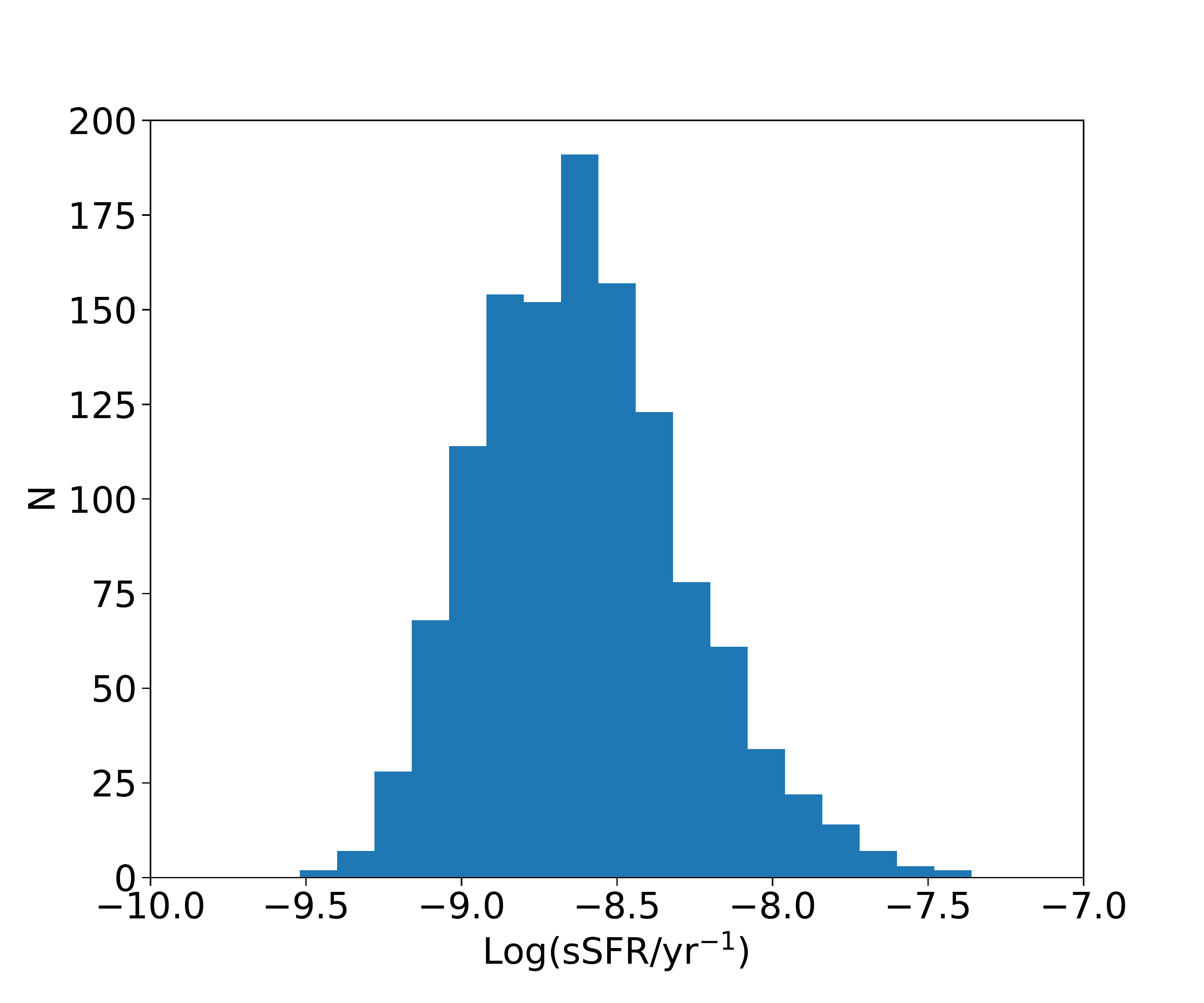}
        \hspace{-0.4cm}
		\includegraphics[width=0.5\columnwidth]{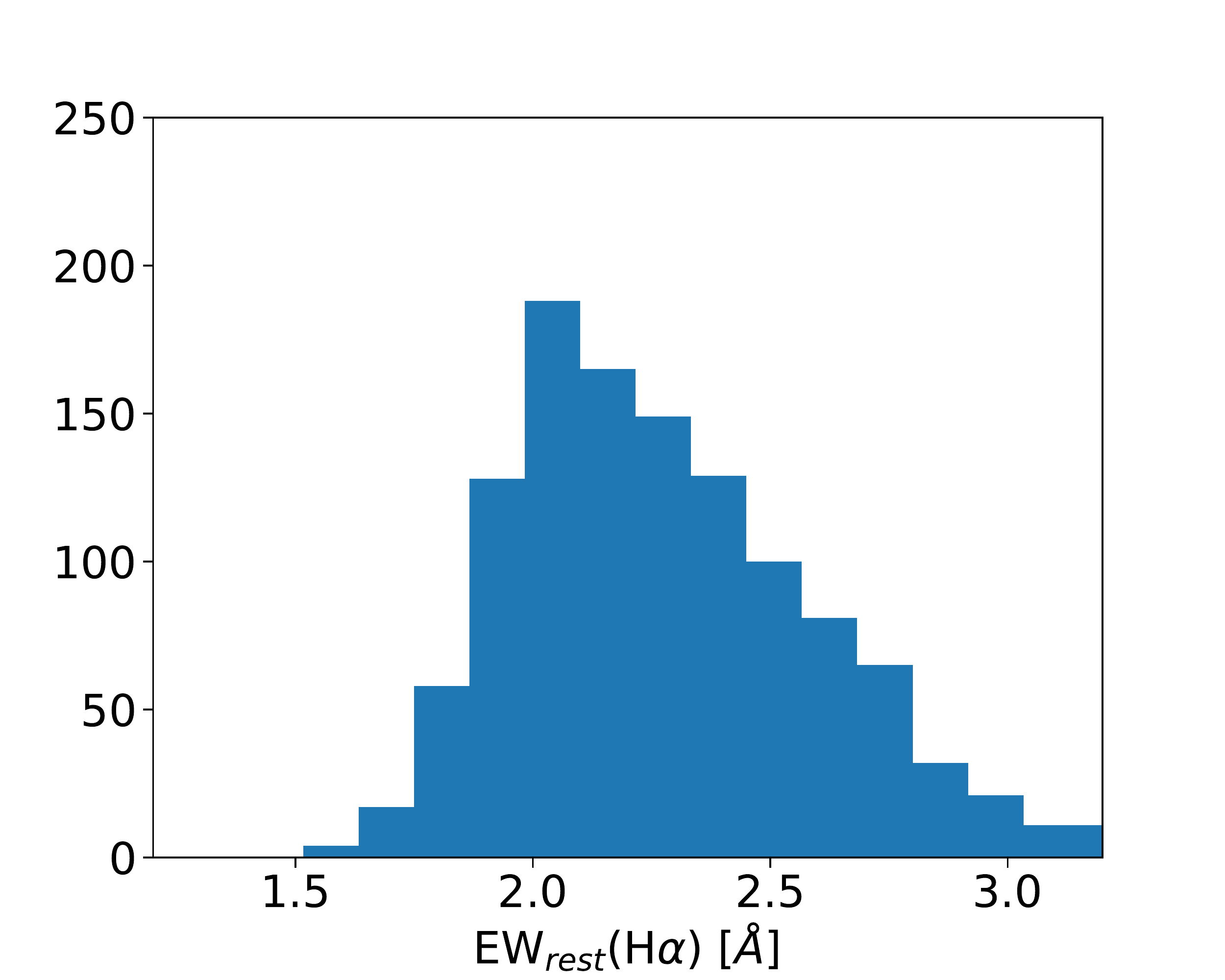}
    \caption{{\bf Left:} specific star formation rate distribution from our sample, computed from the \ha\ emission following \citet{kennicutt12} calibration and corrected for dust attenuation using the \citet{cardelli89} law. {\bf Right:} distribution of the \ha\ equivalent width calculated in the rest frame.}
    \label{fig:histo}
\end{figure}

\section{Star Formation Burstiness in low-mass galaxies}
\label{sec:burstiness}
While star formation is expected to proceed smoothly in massive galaxies, it becomes highly stochastic in lower mass galaxies, with short-lived intense bursts of star formation. These bursts are marked by high \ha\ equivalent but also high \xiion\ as shown by hydrodynamical simulations \citep{shen14, dominguez15}. In this section, we will explore the effects of star formation burstiness on the galaxy observables.

\subsection[SFR-Mass relation]{SFR-\mstar\ relation}

Observations from the local universe to high-redshift galaxies show a tight correlation between the star formation rate and stellar mass \citep[e.g.][]{brinchmann04,noeske07,whitaker14}. The underlying picture is that galaxies accrete gas and build up their stellar mass in a gradual and smooth process, with constant SFR averaged over hundreds of Myrs. However, during the last decade, observations of low-mass galaxies, primarily selected upon their strong emission lines, have painted a different picture of star formation processes. In particular, extreme emission line galaxies (EELGs) experience short episodes of intense star formation, leading to excursions from the SFR-\mstar\ main sequence \citep{weisz12,atek14c,guo16,faisst19,emami19}, although, the averaged SFR over cosmic time, would match a steady star formation scenario. Several hydrodynamical simulations also unveiled the stochastic processes at play in low-mass galaxies compared to their massive counterparts \citep{shen14,dominguez15,sparre17}, as a result of episodes of gas accretion that trigger star formation followed by stellar feedback and gas outflows, which hampers gas infall into the galaxy.

\begin{figure}
    \centering
    \includegraphics[width=8.7cm]{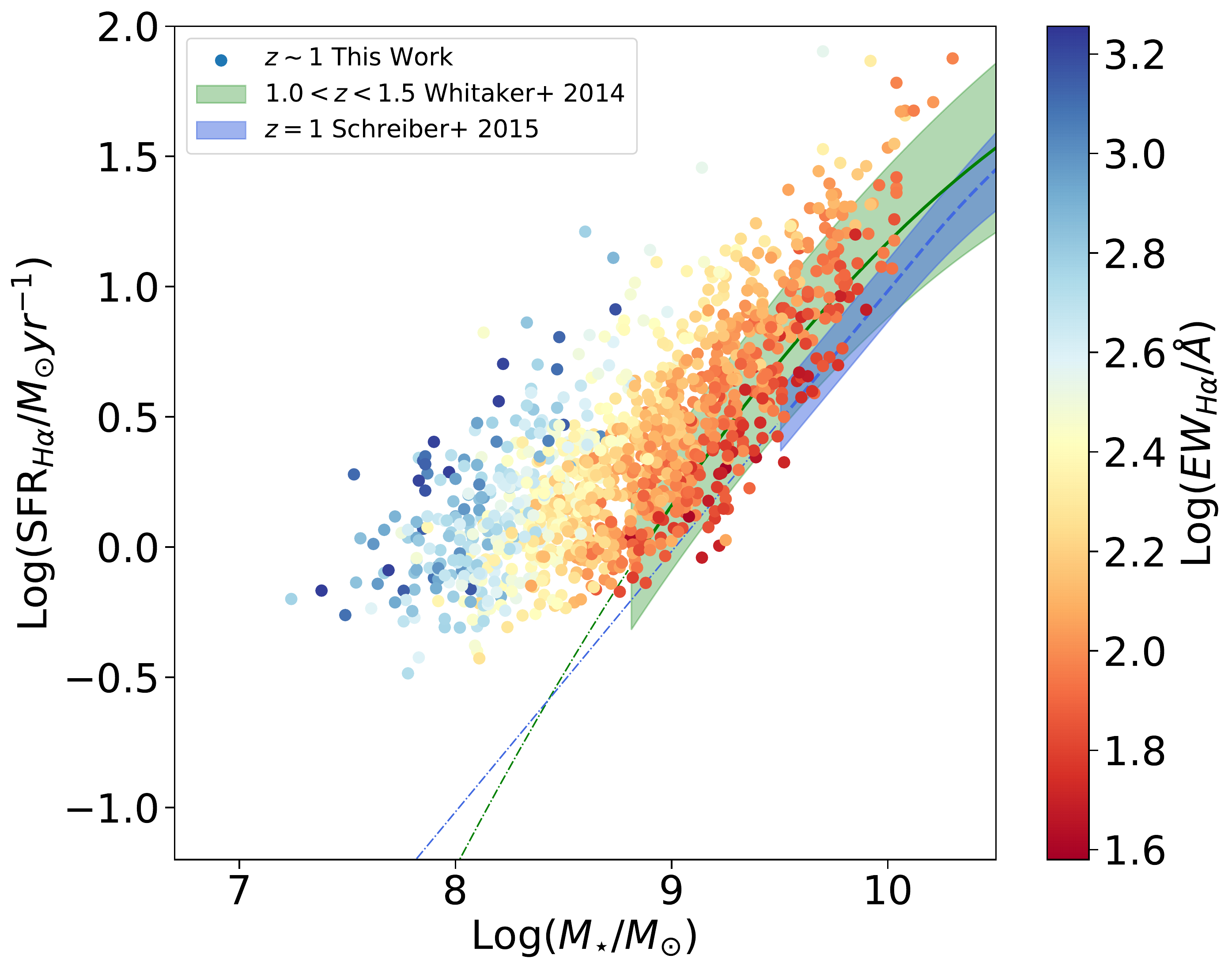}
    \caption{The star formation rate as function of stellar mass, known as the star formation main sequence. SFR(\ha) is corrected for dust attenuation using the Cardelli law. Our data points are also colored according to their \ha\ equivalent width. The green solid line and the blue dashed line show the results of \citet{whitaker14} for $1<z<1.5$ galaxies and the relation derived by \citet{schreiber15} at $z=1$, respectively. the shaded areas are the associated uncertainties, while the dot-dashed lines represent an extrapolation of the above-mentioned relations below their stellar mass limits.
    }
    \label{fig:sfr_mass}
\end{figure}

We first use the \ha\ emission to compute the SFR following the \citet{kennicutt12} calibration, assuming a solar metallicity and a \citet{chabrier03} IMF, and a Cardelli dust attenuation correction. The SED fitting procedure used to constrain the stellar mass also assumes a Chabrier IMF \citep{skelton14}. In Figure \ref{fig:sfr_mass}, we plot the correlation between the SFR and \mstar\ for our sample of galaxies together with the best-fit relations in the literature at similar redshifts, which combined the observed UV and reprocessed IR light to derive the total SFR. \citet{whitaker14} derived the total IR luminosity using a luminosity-independent conversion factor from the {\em Spitzer}/MIPS 24 $\mu$m based on a single SED template \citep{wuyts11b}. \citet{schreiber15} compute the total IR luminosity by fitting SED templates from \citet{chary01} to the {\em Herschel} flux densities and measure L$_{IR}$ in the best-fit template. A significant fraction of the sample lies above the main sequence relations. We can see that the departure from the main sequence is associated with an increase in \ewha. In fact, only galaxies with \ewha\ around a hundred \AA\ follow the main sequence. AS explained earlier, this is the result of the elevated instantaneous SFR, probed by \ha, following a recent burst of star formation. The observed flattening of the slope towards low-mass galaxies is likely the result of \ha\ flux incompleteness, as we reach SFRs below 1 \msolyr. However, it is clear that the deviation from the main sequence is much larger at low mass, as the highest \ewha\ values are observed at lower masses. 

\begin{figure}
    \centering
    \includegraphics[width=8.7cm]{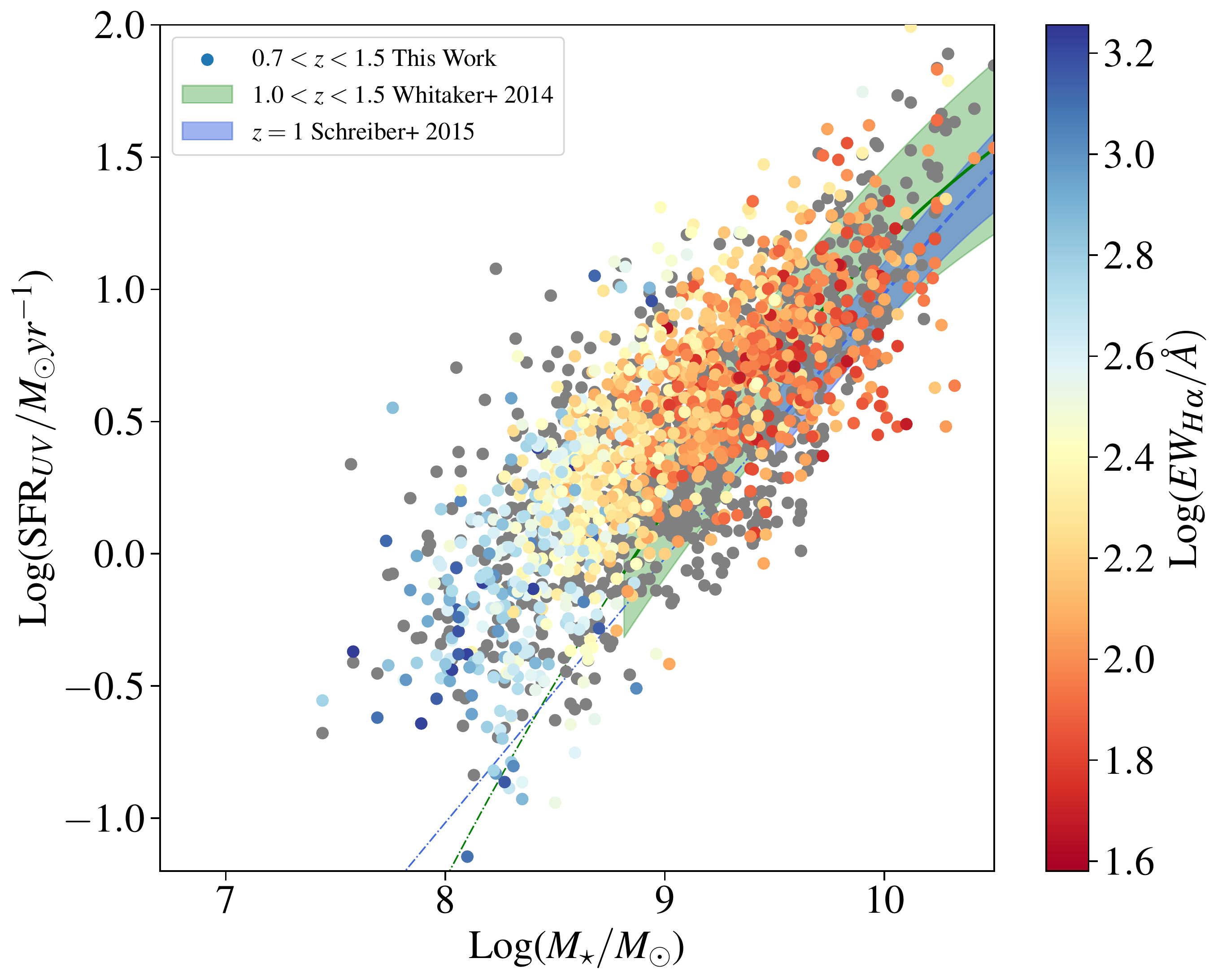}
    \caption{The star formation rate as derived from the UV (\sfruv) as a function of the stellar mass. The colored points are the same as in Figure \ref{fig:sfr_mass}, except the SFR is derived from the UV emission and corrected for dust extinction using the SMC law. In addition, the gray points represent the total SFR, which is the sum of the uncorrected \sfruv\ and the \sfrir\ \citep{skelton14}.
    }
    \label{fig:sfruv_mass}
\end{figure}

\begin{figure}
    \centering
    \includegraphics[width=8.2cm]{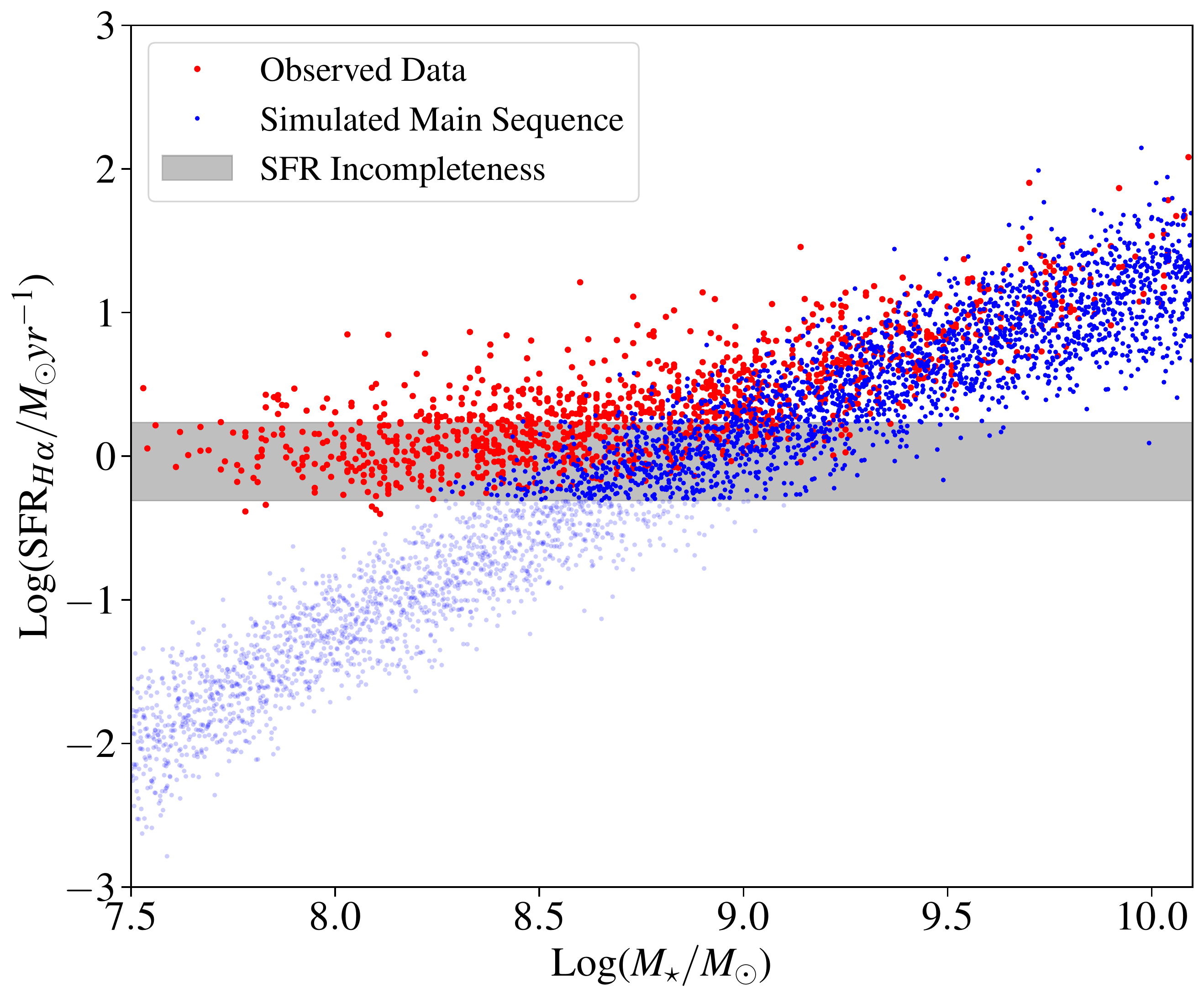}
    \caption{The evolution of the scatter in the SFR-M relation. The blue points represent simulated galaxies following the Main Sequence relation and scatter of \citet{whitaker14}. The light-blue points are the galaxies excluded by the \ha\ flux selection used in our observations. The gray shaded area represents the \sfrha\ limit in the redshift range of our sample. The red points are the observed galaxies shown in Fig. \ref{fig:sfr_mass}.  
    }
    \label{fig:sfr_mass_sim}
\end{figure}

Regardless of the SFR limit, the scatter seems to increase towards lower masses, indicating a higher stochasticity in star formation histories. In order to quantify the apparent increase of the scatter towards lower-galaxies, we populated the SFR-\mstar\ plane with simulated galaxies following the MS relation of \citet{whitaker14} with the corresponding scatter. Then we applied a similar \ha\ flux selection used in the observations. We can see the results in Fig. \ref{fig:sfr_mass_sim}. The blue points follow the expected MS trend, but below a stellar mass of Log(\mstar/\msol)$\sim 9$, the \ha\ flux selection removes most of the galaxies (represented by light blue points). The corresponding completeness limit in SFR(\ha) is indicated by the gray shaded region. In our observations, we are still detecting a significant fraction of galaxies (red points) below Log(\mstar/\msol)$\sim 9$, most probably because of an increased scatter towards lower masses. A good agreement between the simulations and observations in the mass range below Log(\mstar/\msol)$\sim 9$ is achieved by using a scatter a factor of 3 higher than the one observed at higher masses. This stands in contrast with other studies based on UV-selected samples or SED-based SFR indicators, who found no evidence of increasing scatter towards lower masses \citep{kurczynski16}, nor evidence for an enhanced \ha\ luminosity compared to UV \citep{guo16}. \citet{boogaard18} also explored the low-mass end of the main sequence at $0.1<z<0.9$  by using Balmer lines to measure the SFR, finding an intrinsic scatter similar to \citet{kurczynski16} and a shallower slope compared to massive galaxy samples.

In Figure \ref{fig:sfruv_mass}, we show how our sample fits in the main sequence diagram when the SFR is derived from the UV. When compared to Figure \ref{fig:sfr_mass}, the galaxies follow more closely the MS, although these relations are extrapolated here (dashed lines) at lower masses (mainly below $10^{9}$ \msol). This follows the scenario where, recent star formation traced by \ha, likely causes the excursion of galaxies out of the MS. In the same figure, we also compare the dust-corrected \sfruv\ with the total SFR (represented by gray points), which was estimated by adding up the uncorrected \sfruv\ and the \sfrir\ derived from the IR  luminosity at $8-1000$ \mic\ measured from {\em Spitzer}/MIPS 24 \mic\ photometry \citep{skelton14}. Although about 20\% of the sample is missing because they are not detected in the {\em Spitzer} data, we can see the dust corrected SFR is globally in fair agreement with the total SFR.

\subsection[Halpha and UV SFR indicators]{\ha\ and UV SFR indicators}
\label{sec:sfrs_m}
The ratio between \sfrha\ and \sfruv\ is often used an indicator of star formation burstiness in galaxies \citep[e.g.,][]{meurer09, weisz12, guo16, broussard19}. At the onset of a burst, the \ha\ flux increases over a very short timescale of $\sim 10$ Myr, whereas UV bright and less massive stars will run on a longer period  $\sim 100$ Myr) \citep[e.g.][]{leitherer99, hao11, kennicutt12, calzetti13}. Therefore, important variations in \sfrha/\sfruv\ ratio are expected when the star formation is non-steady and out of equilibrium. For instance, \citet{weisz12} produced a set of SFH history models to match the observations of a sample of local galaxies. They find that low-mass galaxies experience a bursty star formation, with burst episodes lasting less than 10 Myr and a period of 250 Myr between bursts. Those models appear to reproduce the observed variations of the $f(H\alpha)/f(UV)$ ratio, which are mostly below unity. \citet{emami19} improved this analysis with a larger parameter space with exponentially rising and declining SFH and find similar results, albeit with a larger burst duration. In their analysis, they explore the star formation burstiness by comparing the $L(H\alpha)/L(UV)$ with $\Delta L(H\alpha)$\footnote{$\Delta L(H\alpha)$ is the deviation of the \ha\ luminosity from the best fit to the main sequence of $L(H\alpha)$ {\em vs} \mstar.}, which are correlated, in particular at lower stellar masses. The bursty nature of star formation in low-mass galaxies has been confirmed by hydrodynamical simulations. In the FIRE simulations, \citet{sparre17} show that the SFR probed on a short timescale by \ha\ can vary by an order of magnitude compared to the UV indicator. They also find a significant fraction of galaxies with \sfrha/\sfruv\ ratios well below unity.   

In general, the explanation for the decline of the \sfrha/\sfruv\ ratio observed in those studies is based on the different timescales of these emissions. At the onset of the burst, the \ha\ emission is brighter than the FUV. The stars responsible for the FUV emission have a longer lifetime. This means that the FUV emission will gradually take over, and the \sfrha/\sfruv\ will decline steadily below unity over the course of the burst in the first $\sim 50$ Myr. After the end of the burst, the \ha\ emission from massive stars becomes negligible, as the stars responsible for this emission die after $\sim 10$ Myr. Over the next $100-200$ Myr, the FUV emission in turn starts to fade, as the less massive stars start to disappear, leading to a steady increase in the \sfrha/\sfruv\ ratio. Around $300$ Myr, the \sfrha/\sfruv\ returns to its equilibrium value of unity, which marks the end of the burst cycle \citep[][]{weisz12, dominguez15, emami19}. According to this scenario, the observed \sfrha/\sfruv\ ratio is below unity during most of the burst cycle.

\begin{figure}
        \includegraphics[width=9cm]{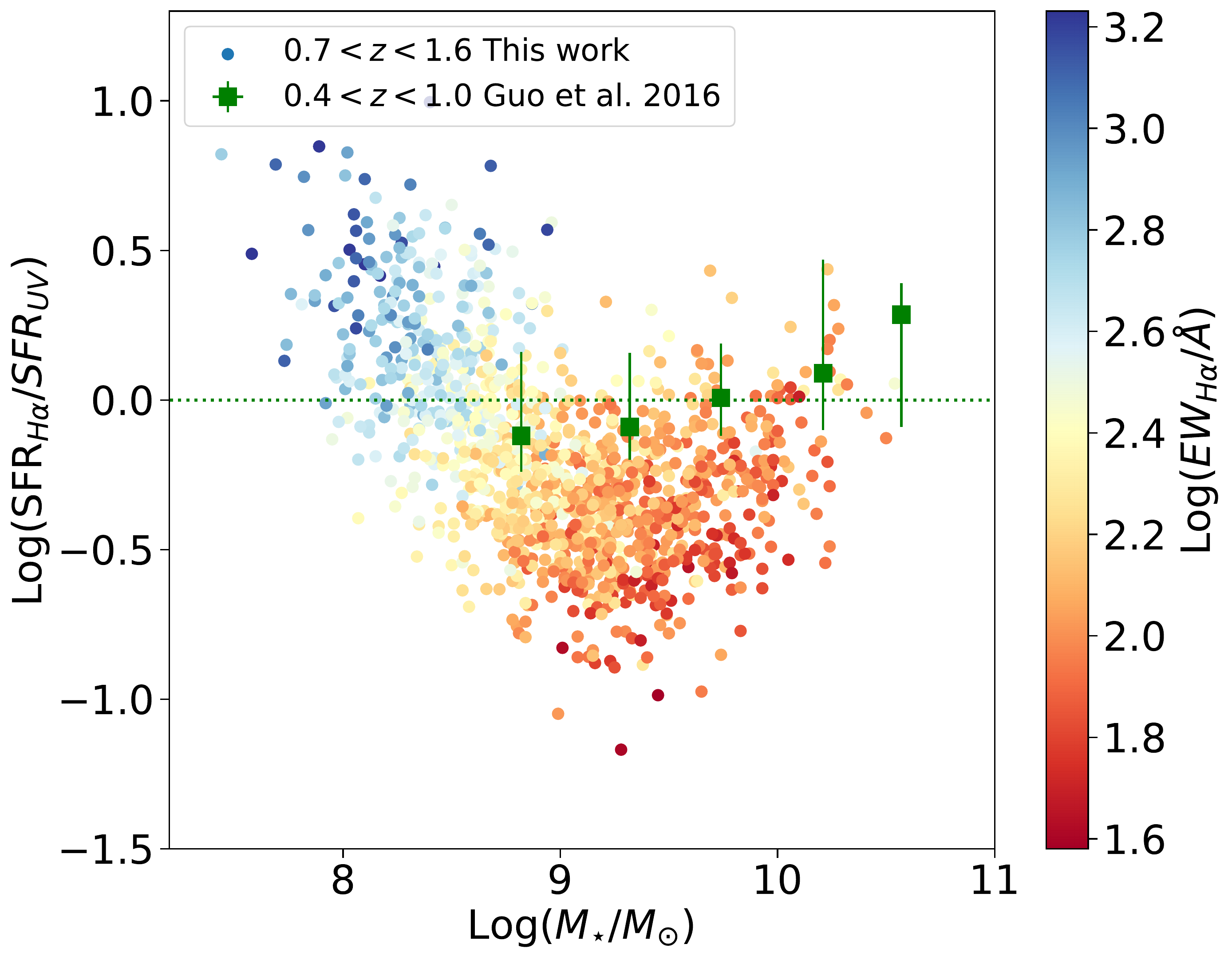}
    \caption{The ratio of \ha\ and UV SFR indicators as a function of stellar mass. \sfrha\ is corrected for dust attenuation using the Cardelli law and \sfruv\ with the SMC curve. Data points are also color-coded with \ewha. The horizontal line indicates a one-to-one ratio. The results of \citet{guo16} are show with green squares. Note that these points represent the ratio SFR$_{H\beta}$/\sfruv.}
    \label{fig:sfrs_m}
\end{figure}

In Figure \ref{fig:sfrs_m}, we plot the \sfrha/\sfruv\ ratio as a function of stellar mass for our sample, assuming an SMC dust correction for the UV continuum. At high masses ($>10^{9}$\msol), the ratio of SFR indicators is mainly close or below unity. For galaxies with \mstar\ $<10^{9}$\msol, the ratio is around unity or higher. From the same figure, we can see a similar trend with \ewha, as galaxies with \ewha\ of a few hundreds to a thousand \AA\ are the ones with higher SFR ratios. 
The main result is that we observe an \sfrha/\sfruv\ ratio above unity mainly in low-mass galaxies (\mstar\ $<10^{9}$\msol) rather than in more massive galaxies (\mstar\ $>10^{9}$\msol). This result is different from what is observed in the studies mentioned earlier and the associated scenario, where \sfrha/\sfruv\ is observed to be mainly below unity. Of course, because of our \ha\ flux selection (see Section \ref{sec:sample} and Figure \ref{fig:sfr_mass_sim}), we might be missing \sfrha/\sfruv\ lower than unity at \mstar\ $<10^{9}$\msol. However, the observation of elevated \sfrha/\sfruv\ in low-mass galaxies still stands, and is not observed in more massive galaxies.

For comparison, the other notable studies along these findings are \citet{shim11} and \citet{faisst19} who analyze a sample of $z \sim 4$ galaxies for which they measure \ha\ through the IRAC 3.6\mic\ excess. \citet{shim11} find an SFR ratio of \sfrha/\sfruv\ $=6.1 \pm 4.1$, even higher than our result. Not surprisingly, their sample also has a very high \ha\ equivalent width distribution (\ewha = 140-1700 \AA), due to the IRAC selection, which is sensitive to \ha\ emitters with \ewha\ higher than 70-350 \AA. This effect is clearly visible in Fig. \ref{fig:sfrs_m}, where high-\ewha\ galaxies have the highest \sfrha/\sfruv. Similarly, \citet{faisst19} find that more than 50\% of their sample has \sfrha\ in excess compared to \sfruv, particularly in low-mass galaxies. They also find an anti-correlation between \ewha\ and stellar mass. The trend and SFR ratios observed at low masses differ from the results of some previous studies summarized above \citep{weisz12,sparre17}. This was already observed in Figure \ref{fig:sfr_mass} where the scatter of the main sequence (i.e. $\Delta L(H\alpha)$) is larger at lower masses. We note that \citet{emami19} also find a larger scatter int the L(H$\alpha$) vs \mstar\ relation towards lower mass galaxies.

Given the timescales traced by \ha\ and UV, the distribution of \sfrha/\sfruv\ and its dispersion provides an indication on the SFH parameters. As discussed earlier, since \sfruv\ runs on hundreds Myr, while \sfrha\ on 10 Myr, the distribution of observed galaxies should be centered on the low side of \sfrha/\sfruv. However, a variation in the following two parameters will increase the prevalence of galaxies with higher SFR ratios like the ones observed here: (i) a larger $\tau$ in the exponential SF; or (ii) a higher duty cycle, i.e. a shorter period between successive SF bursts.
 
In this context, the results of \citet{guo16} are also shown in Fig. \ref{fig:sfr_mass}. They selected a sample of 165 galaxies at $0.4<z<1$ with SNR(\hb)$>3$ and SNR(F275W)$>5$. They derived \sfrha\ using the \hb\ flux, the case B recombination ratio of 2.86, and the calibration of \citet{kennicutt12}. They derived \sfruv\ using the UV luminosity at 1500 \AA\ and the same calibration of \citet{kennicutt12}. Both determinations assume a Chabrier IMF and solar metallicity. We used the same calibration and parameters to derive the SFRs of the present sample. \citet{guo16} also showed that dust correction is negligible for these SFR determinations and apply a correction only for galaxies with SFR$_{\mathrm{UV + IR}}$/SFR$_{1500A} > 20$. They do not find the mass-dependant trend observed in our sample, but rather an increase in the SFR ratio with increasing mass. Again, it is possible that these differences are partially due to selection effects, since their sample has a median $EW_{H\beta} = 15$ \AA, a factor of two lower than our sample. We argue that if this were only due to selection effects, these high-EW lines should also be present in higher-mass systems. Our sample most likely probes a low-mass galaxy population caught during a burst of star formation, rather than its decline as seen in those studies. Theoretical models of bursty star formation also predict an increase of this mode in low-mass galaxies \citep[e.g.][]{faucher18} due to shorter dynamical timescales and probably higher gas fractions.

Another possible source for the variations in the \sfrha/\sfruv\ ratio in low-mass galaxies is a stochastic sampling of the IMF. In the case of low-mass galaxies with a low SFR, massive stars could form less frequently than assumed, leading to deviations of \sfrha/\sfruv\ from equilibrium even in the case of a steady star formation. Numerous studies have explored the effects on IMF sampling \citep{pflamm09,fumagalli11, weisz12, guo16, emami19}. Overall, the resulting variations in SFR indicators were insufficient to reproduce the observed decline of \sfrha/\sfruv. The discrepancy is even larger for our results, where we observe a clear increase towards lower masses, which makes the IMF sampling an unlikely scenario. 

Finally, several uncertainties remain regarding the determination of the \sfrha/\sfruv\ ratio and its evolution as a function of galaxy parameters. Both SFR indicators use conversion factors from \ha\ and UV luminosities assuming a constant star formation. Of course, we expect important deviations from these assumptions for different SFHs, particularly in the case of bursty star formation. If both SFR calibrations were correct, we would expect the \sfrha\ and \sfruv\ to be equal. This is precisely why such a ratio can reveal deviations from steady star formation scenarios and trace bursty SFH. In addition to SFH, the inclusion of binary stars in stellar population models and a higher mass end (300 \msol) for the IMF \citep{wilkins19} can increase \sfruv\ by 0.15-0.2 dex, and \sfrha\ by 0.35 dex. \citet{castellano14} also showed that \sfruv\ tend to underestimate the true SFR of $z \sim 3$ Lyman break galaxies with sub-solar metallicity by a factor of a few. It is unclear whether this underestimate would also apply to lower-mass galaxies at lower redshifts. Furthermore, if the attenuation curve for the stellar continuum is different in lower-mass galaxies \citep[e.g.,][]{salim20}, it will also affect the SFR ratio. One scenario that would contribute to the increase of \sfrha/\sfruv\ ratio towards lower mass galaxies is the one where the dust attenuation law becomes steeper towards low-mass galaxies.

\section{ionizing efficiency}
\label{sec:xi}

We compute the ionizing efficiency \xiion, which is defined as the ratio between the ionizing (Lyman continuum) photon production \nion\ and the observed non-ionizing UV luminosity density \luv\ estimated at 1500 \AA: 

\begin{equation}\label{eq:xi}
    \xi_{\rm ion} = \frac{\dot{N}_{\rm ion}} {L_{UV}} ~[erg^{-1}~ Hz]
\end{equation}

In the framework of the case B recombination theory \citep{osterbrock89}, $\dot{N}_{\rm ion}$ can be estimated from Hydrogen recombination lines \citep{leitherer95}: 
\begin{equation}\label{eq:nh}
    L(H\alpha)~[erg ~s^{-1}] = 1.36 \times (1-f_{esc})~ 10^{-12} ~{\dot{N}_{\rm ion}} ~[s^{-1}]  
\end{equation}
where $ L(H\alpha)$ is the \ha\ luminosity and \fesc\ is the escape fraction of Lyman continuum radiation from galaxies. Most results in the literature indicate a wide range of \fesc\ values, both in low- and intermediate- redshift galaxies. Recent spectroscopic measurements in $z \sim 0.3$ compact star-forming galaxies reported by \citet{izotov16,izotov18} indicate an escape fraction between 2 and 70\%. Imaging campaigns at higher redshifts reported a few robust detections with escape fraction exceeding \fesc\ $\sim 50\%$ \citep[e.g.,][]{shapley16,vanzella16,bian17,vanzella18,rivera19}. On the other hand, \citet{rutkowski17} estimate an upper limit of \fesc\ $<15$ \% in emission-line galaxies at $z \sim 2.5$, using partly the 3DHST G141-grism data in GOODS-N and UVUDF fields. For the sake of consistency with literature results, here we assume a zero escape fraction of LyC radiation. Accounting for \fesc(LyC) will result in a higher \xiion. 

According to equations \ref{eq:xi} and \ref{eq:nh}, \xiion\ is equivalent to the ratio \sfrha/\sfruv, modulo a factor of $1.3 \times 10^{26} ~ (1-f_{esc})$. Therefore, in the case where we assume a zero escape fraction, this burstiness indicator is a direct probe of the ionizing efficiency and its evolution with galaxy parameters. The ionizing efficiency will be subject to the same uncertainties discussed earlier for the SFR ratio. 

\begin{figure}
        \includegraphics[width=\columnwidth]{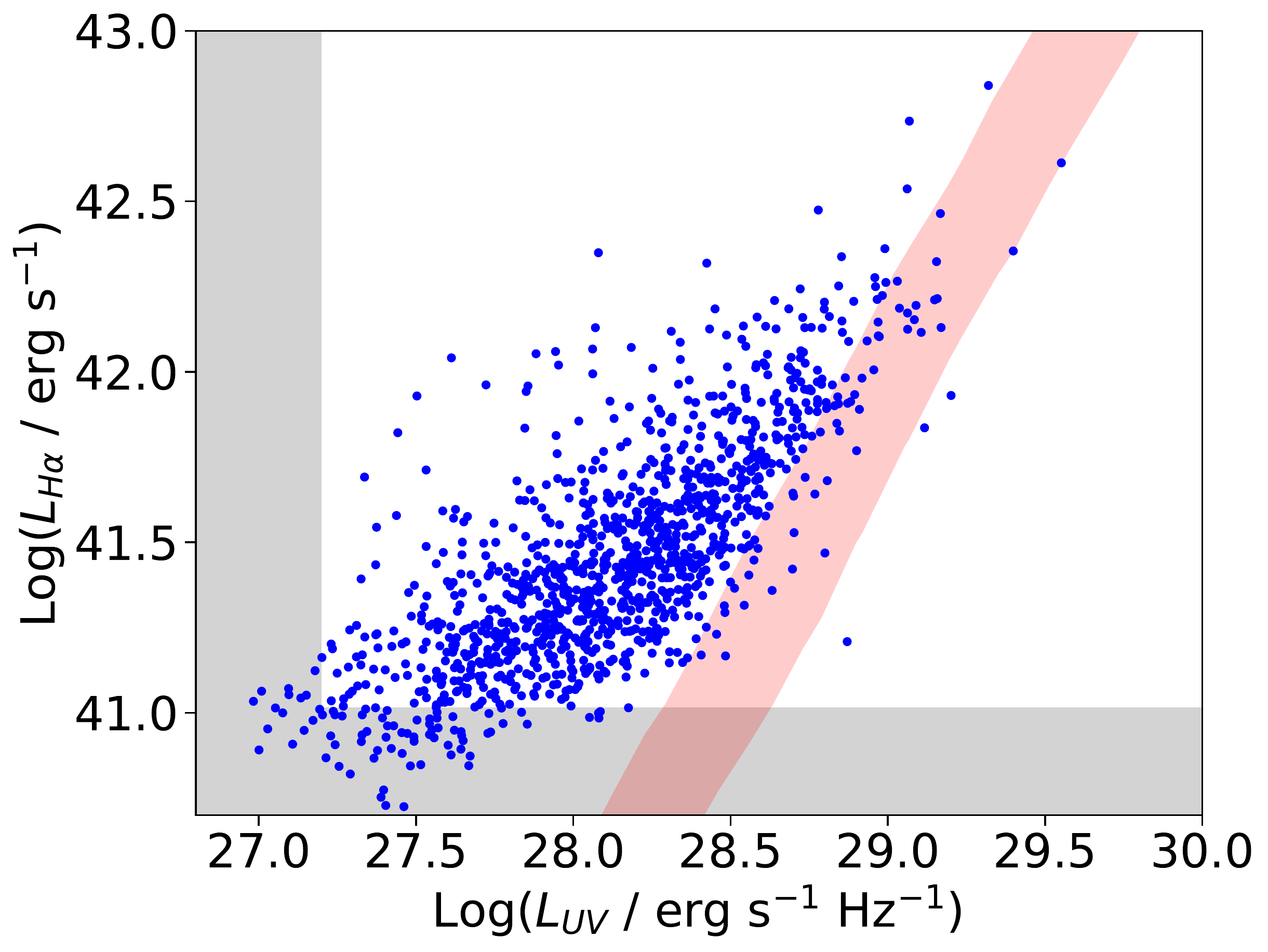}
    \caption{The \ha\ luminosity as a function of rest-frame UV luminosity for the \xiion\ sample. The red-shaded region represents the \xiion\ lower limit imposed by our equivalent width cut (cf. Figure \ref{fig:comp}). The associated range of values account for the redshift range $0.7<z<1.5$ and the relationship between the IR and UV magnitudes (cf. text for details). Shaded regions along both axes indicate the detection limits in L(\ha) and L(UV), calculated at the mean redshift of the sample.}
    \label{fig:lha_luv}
\end{figure}

In Figure \ref{fig:lha_luv}, we demonstrate how our sample selection translates into \xiion\ completeness. We plot the \ha\ luminosity as a function of UV luminosity for our \xiion\ sample, together with their respective limits (gray shaded regions) derived from the depth of the observations. Combining the \ewha\ cut and \ha\ flux limit, and using the observed correlation between the UV and IR magnitude, we computed the \xiion\ limit (red-shaded area), assuming a redshift range of $0.7<z<1.5$ in the simulations (cf. Appendix \ref{app:a1}). Overall, these simulations will be helpful in assessing the limits of the correlations we explore in Section \ref{sec:prop}.

\subsection{Ionizing efficiency results}
\label{sec:results}
\begin{figure}
        \includegraphics[width=\columnwidth]{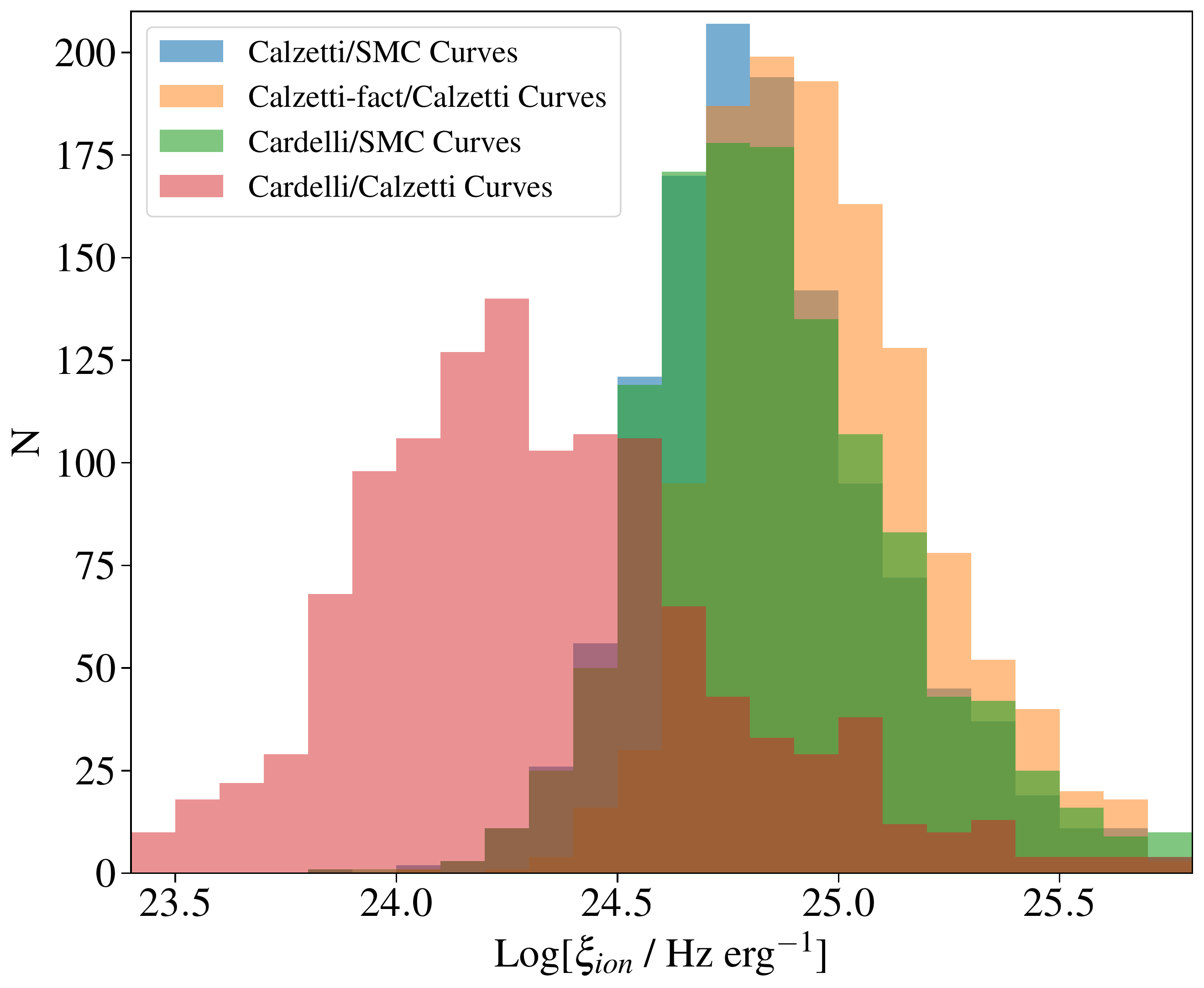}
    \caption{Distribution of the ionizing efficiency \xiion. The blue, orange, green, and red histograms show the dust-corrected measurements using three different combinations of attenuation curves for the nebular emission and the UV continuum: \citet{cardelli89}/SMC,  \citet{calzetti00}/\citet{calzetti00} but assuming the nebular attenuation is a factor of (1/0.44) higher than the stellar attenuation, \citet{calzetti00}/SMC, and \citet{cardelli89}/\citet{calzetti00} curves, respectively. The main combination used in Section \ref{sec:burstiness} yields the green distribution.
    The median values for each configuration are Log(\xiion/erg$^{-1}$ Hz)$=24.80 \pm 0.26$, $24.93 \pm 0.24$, $24.83\pm 0.27$, $24.26 \pm 0.36$.}
    \label{fig:histo_xi}
\end{figure}

\begin{figure}
    \centering
    \includegraphics[width=\columnwidth]{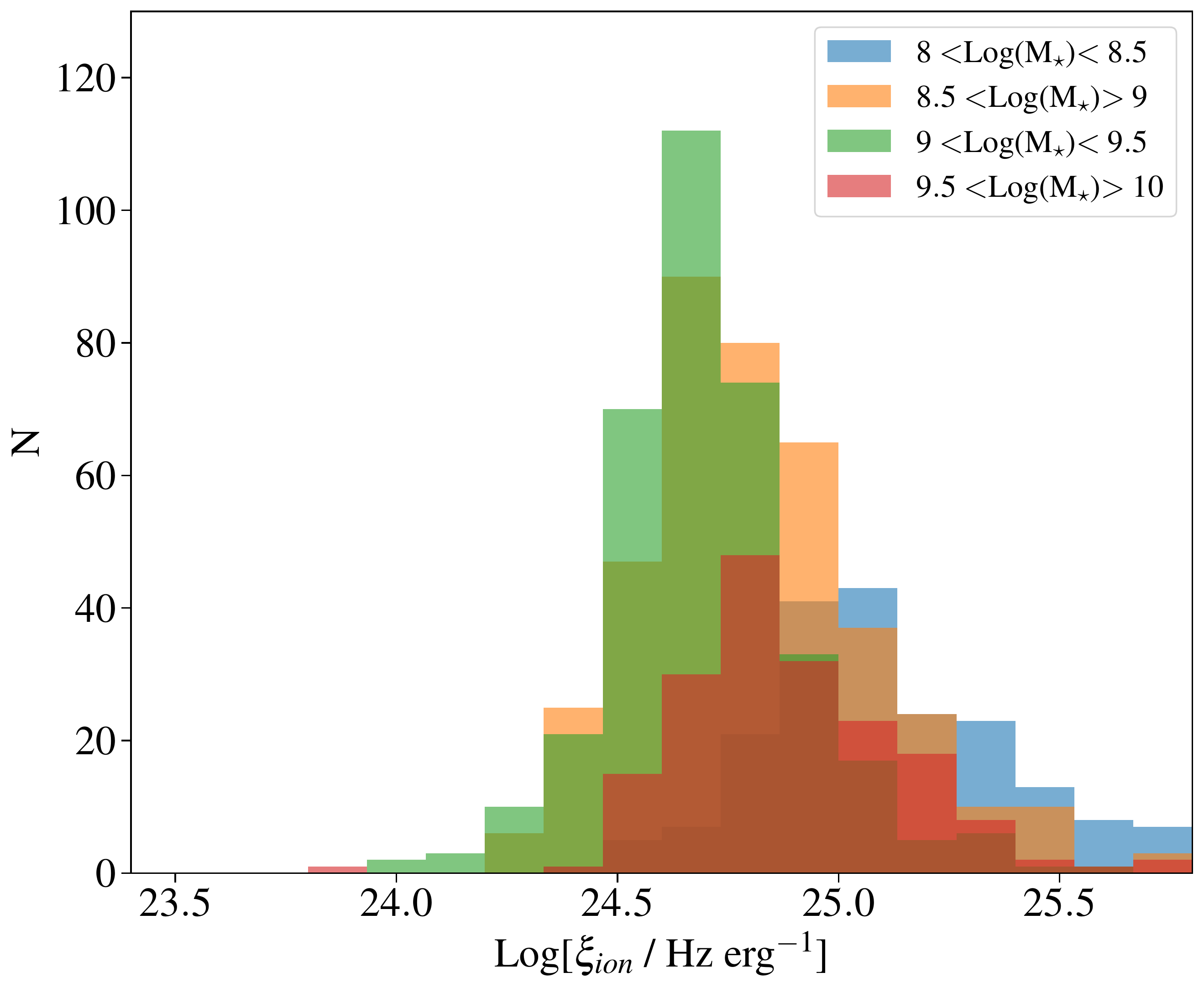}
    \caption{The \xiion\ distribution in four different bins of stellar mass. The Cardelli/SMC combination is adopted for dust correction.}
    \label{fig:histo_xi_mbin}
\end{figure}

The results of the dust-corrected estimates of \xiion\ are shown in Figure \ref{fig:histo_xi}. Adopting standard attenuation corrections using the \citet{cardelli89}  law for \ha\ emission and the SMC curve for UV continuum, we obtain a median ionizing efficiency of Log(\xiion/erg$^{-1}$ Hz)$=24.80 \pm 0.26$. Adopting the Calzetti law for the nebular attenuation has little effect on the ionizing efficiency, except when deriving the nebular attenuation by applying a factor of 1/0.44 to the stellar attenuation \citep{calzetti00}. The \xiion\ estimates are more sensitive to the attenuation curve in the UV domain. The derived \xiion\ are lower by 0.45 dex if we assume the Calzetti attenuation curve rather than the SMC curve for dust-correcting the UV luminosity density. The scatter in the \xiion\ distribution increases by 0.11 dex in this case. Table \ref{tab:dust} summarizes the relative importance of dust correction to the \ha\ and UV emissions. We can see that the correction factor applied to the UV emission is much larger than the one for \ha. Also, the choice of an attenuation law is more critical for the UV emission. In figure \ref{fig:histo_xi_mbin}, we show the \xiion\ distribution in four mass bins to highlight the dependence on stellar mass discussed later in Section \ref{sec:prop}.

These values are in general significantly lower than what has been found in previous studies and the canonical value of Log(\xiion)$=25.2$ widely used in computing the ionizing budget of galaxies \citep[e.g.,][]{robertson13}. As a comparison, \citet{emami20} report an average Log(\xiion)$=25.47 \pm 0.09$ in a sample of lensed galaxies at $z \sim 2$, which is almost an order of magnitude higher. At a similar redshift, \citet{shivaei18} find Log(\xiion)$=25.34$. Both values are derived with an SMC extinction correction of the UV luminosity. 

We list below a literature compilation of \xiion\ determinations at different redshifts.

\begin{itemize}
    \item The results of \citet{emami20} are based on spectroscopic observations of 28 $z \sim 2$ lensed galaxies, with a similar stellar mass range probed by our sample ($8 \leq Log(M_{\star}/M_{\odot}) \leq 10$). The study follows the same approach used here for \xiion\ estimates and dust correction. However, their results are based on stacked spectra, in bins of physical properties (stellar mass, UV magnitude, UV continuum slope). 
    
    \item \citet{shivaei18} used spectroscopic and UV imaging data from the MOSDEF survey to measure \xiion\ in a large sample of 673 galaxies at $1.4 < z < 2.6$. They use the Balmer decrement \ha/\hb\ to measure nebular attenuation, whereas it is available for only a sub-sample of our galaxies (cf. Section \ref{sec:dust}). Stellar attenuation is derived from SED fitting. Importantly, that work probes stellar masses above $Log(M_{\star}/M_{\odot}) > 9$, which is significantly higher than the present sample. 
    
     \item Using spectroscopic observations to measure \hb\ emission and SED fitting to infer the UV continuum emission, \citet{nakajima16} measured \xiion\ in sample of 15 \lya\ Emitters and Lyman Break Galaxies at $z \sim 3$. No dust attenuation was assumed in this determination. We computed an average \xiion\ in a sub-sample of $Log(M_{\star}/M_{\odot}) \sim 9.7$ galaxies. Their sample also include galaxies with no \hb\ detection and whose \xiion\ upper limits and stellar masses are lower than the average values reported in the Figure.
     
    \item Unlike these studies, \citet{matthee17} used narrowband observations to measure the \ha\ flux and infer \xiion\ in a sample of $z \sim 2.2$ galaxies. We select here their low-mass sample of \ha\ emitters with an average stellar mass of Log($M_{\star}/M_{\odot}) \sim 9.2$).

    \item In addition, several studies have used the broad-band excess between the $3.6\mu m$ and $4.5\mu m$ {\em Spitzer}/IRAC channels to derive the \ha\ flux of $z > 4$ galaxies. Based on this method, \citet{bouwens16} derived \xiion\ for galaxies at $z \sim 4.4$ and $z \sim 5.25$. These values were corrected for dust using a stellar attenuation derived from SED fitting. The results of two attenuation laws (Calzetti and SMC) are shown in the Figure. Their stellar mass range is also comparable to this work, with an average value around Log($M_{\star}/M_{\odot}) \sim 9$).
    
    \item Similarly, \citet{lam19} used stacked IRAC colors to measure \ha\ emission and infer \xiion\ in sample of 300 galaxies at $3.8 < z < 5.3$ with spectroscopic redshifts. The dust correction is based on UV continuum slope measurements and the SMC curve. We here report the average \xiion\ value computed in sub-sample over a stellar mass range of $8 \leq Log(M_{\star}/M_{\odot}) \leq 10$.
    
    \item At higher redshifts, \citet{stark15} obtained rest-frame UV spectroscopy of $z \sim 7$ lensed galaxies and used stellar population and photo-ionization models to infer the ionizing efficiency of one of them. As the authors note, the intense radiation field of this galaxy is not necessarily representative of the galaxy population at $z > 6$ because of the strong equivalent width selection and the bright \lya\ emission. 
\end{itemize}

\begin{figure}
        \includegraphics[width=9cm]{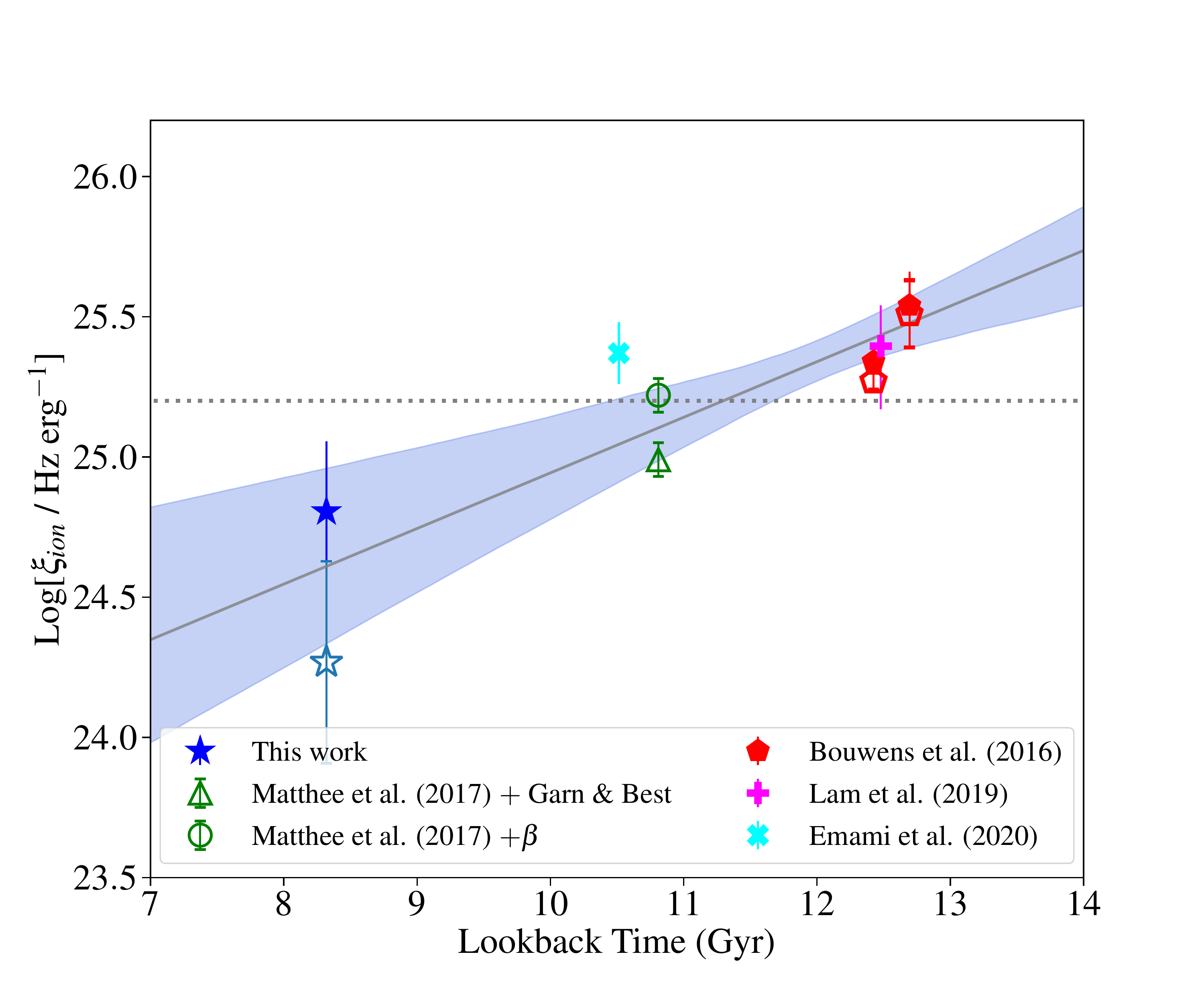}
    \caption{The ionizing efficiency as a function of lookback time. Our measurement at $z \sim 1$ (blue stars) are compared with literature results at higher redshifts. Note that the median stellar mass of our sample is comparable to the other studies on this figure (see text for details). The filled (open) symbols show the difference in \xiion\ measurements when an SMC (Calzetti) curve is used to correct or dust. The dotted horizontal line indicates the canonical value of Log(\xiion)$=25.20$. The gray line and the shaded area show our best-fit results and associated 95\% confidence interval. The fit is performed on all the data points shown here.}
    \label{fig:xi_time}
\end{figure}

Within this list of studies, only \citet{emami20} has both spectroscopic measurements of the \ha\ flux and UV imaging to accurately infer \xiion\ for galaxies in the same range of stellar mass as probed in this study. The main differences are the lower redshift range and the much larger sample size used in the present study. In figure \ref{fig:xi_time}, we keep only the studies that probe a similar stellar mass range and use the same dust prescriptions as our study. We observe a clear evolution of \xiion\ as a function of lookback time $t_{L}$, ranging from Log(\xiion)$=24.70$ at $t_{L} \sim 8.3$ Gyr to Log(\xiion)$=25.54$ at $t_{L} \sim  12.7$ Gyr. We perform a linear fit to all the data points by computing the maximum likelihood using Markov Chain Monte Carlo (MCMC) simulations using {\tt emcee}  \citep{emcee}. In Figure \ref{fig:xi_time}, we plot the best-fit relation and 95\% confidence interval of the fit. We find a best-fit relation of Log(\xiion) $= 0.21 ~ t_{L} + 22.82$. Such evolution is expected since galaxy physical properties are known to evolve with redshift. For instance, the emission-line selection naturally favors high equivalent widths for the \ha\ line and the average \ha\ equivalent width of emission-line galaxies is known to increase with redshift \citep[e.g.,][]{atek11, smit14, faisst16,reddy18}. These effects can explain the higher \xiion\ values observed at higher redshifts (see also Section \ref{sec:prop}).

\section[Evolution of the ionizing efficiency with galaxy properties]{Evolution of \xiion\ with galaxy properties}
\label{sec:prop}
We now explore the evolution of \xiion\ as a function of the galaxy physical properties. 

\subsection[The relation between the ionizing efficiency and emission line equivalent width]{The relation between \xiion\ and \mdseries{\ewha}}

\begin{figure}
        \includegraphics[width=8.5cm]{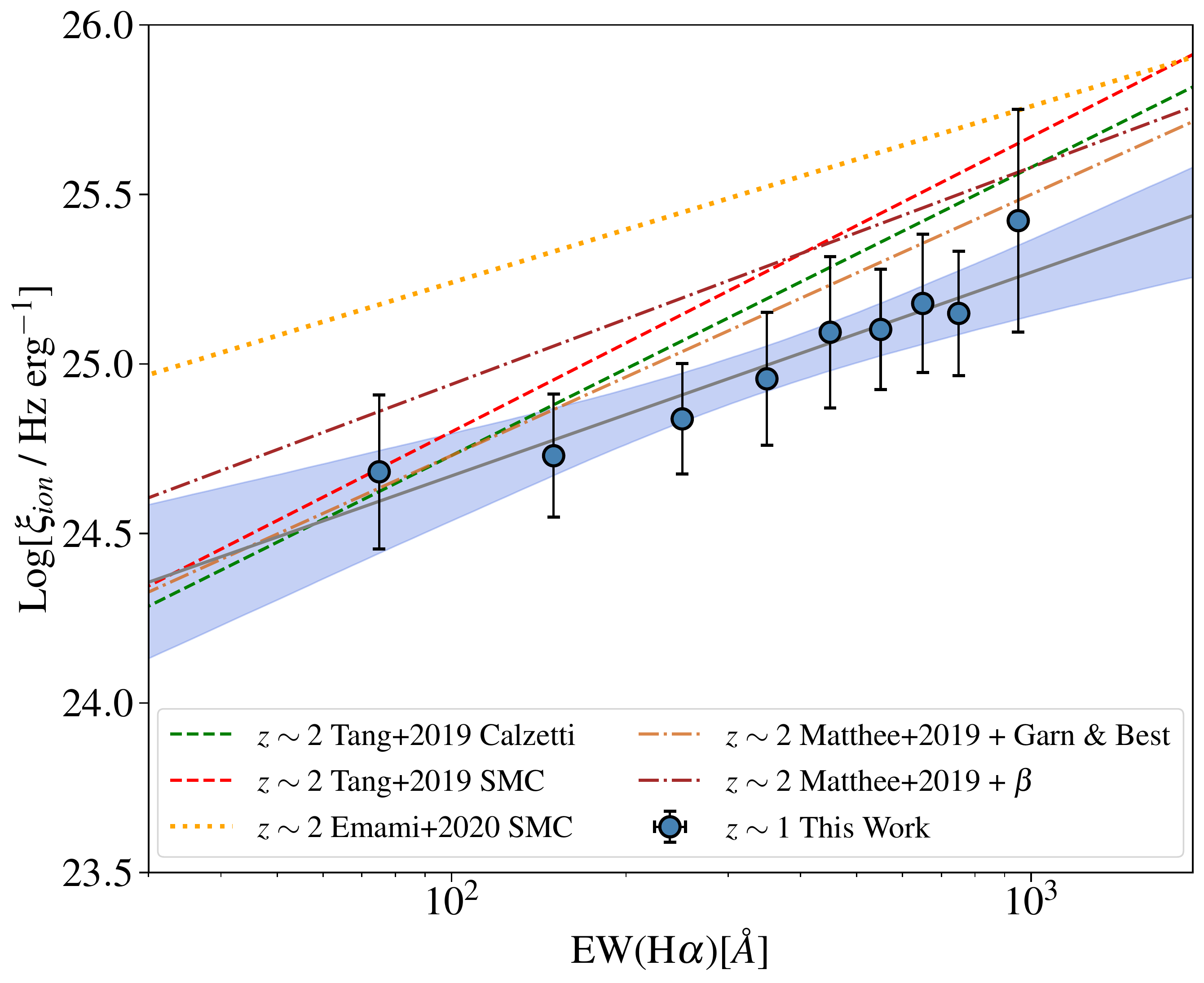}
    \caption{The ionizing efficiency as a function of \ha\ line equivalent width (blue circles). Error bars indicate the dispersion in each \ewha\ bin. The best-fit relation is shown with a gray solid line with associated 68\% confidence interval (shaded area). Other correlations found at higher redshift in the literature are also shown: orange dotted line for \citet{emami20}, green and red dashed lines for \citet{tang19}, orange and brown dot-dashed lines for \citet{matthee17}. The legend also indicates the prescriptions used for dust corrections in each study.}
    \label{fig:xi_ew}
\end{figure}

As previously noted, \xiion\ is directly related to the Balmer recombination lines, but also the UV luminosity of galaxies. It is a measure of the contribution of young massive stars as traced by \ha\ relative to a slightly older stellar population as traced by the non-ionizing stellar continuum. A good proxy for this ratio is the \ha\ equivalent width, which is a good indicator of the age of the stellar population. In Figure \ref{fig:xi_ew}, we show the clear correlation found between \xiion\ and \ewha. The best fit relation is given by:

{\bf
\begin{equation}
 Log(\xi_{ion}) = 0.6^{+0.1}_{-0.1} \times Log(EW_{H\alpha}) + 23.47^{+0.26}_{-0.25}
\end{equation}
}

The fitting procedure is similar to the one used in Figure \ref{fig:xi_time}. On average, the \ha\ equivalent width is well correlated with \xiion\ because it is directly related to the star formation activity and the age of the stellar population. However, part of this tight correlation can also be explained by the fact that the \ha\ luminosity is also used is computing \xiion\ (cf. Equation \ref{eq:nh}, \ref{eq:xi}).

Similarly, this strong relationship between \xiion\ and \ewha\ has been found at other redshifts. Combining a spectroscopic follow-up of strong line emitters and SED fitting, \citet{tang19} finds a similar relation at $z \sim 2$, albeit with a steeper slope of 0.87. We can also appreciate the effect of adopting different prescriptions for the dust correction. This is even more apparent in the case of \citet{matthee17} results, where a significant offset is observed between dust corrections based on the $\beta$ slope or the dust-\mstar\ relation of \citet{garn10}. \citet{emami20} find a shallower slope of 0.53 although with a higher dispersion. Finally, \citep{reddy18b} also explored the relation between \xiion\ and Balmer lines in the MOSDEF survey, over a lightly higher stellar mass range of (\mstar $>10^{9}$ \msol). They find a strong correlation between \xiion\ and \ewha.  

Most of the \xiion\ determinations in the literature are based on samples that preferably select strong emission line galaxies \citep[but see][]{reddy18b}. The extrapolation of these trends to low equivalent widths (below 200\AA) was first questioned by \citet{tang19}. We don't find any significant flattening of the relation towards lower EWs, even though the sample becomes less complete below $EW = 80$ \AA. This result also confirms the findings of \citet{emami20} and \citet{matthee17}.

\subsection[\xiion\ {\em vs} stellar mass and UV magnitude]{Ionizing efficiency {\em vs} stellar mass and UV magnitude}

\begin{figure}
        \includegraphics[width=9cm]{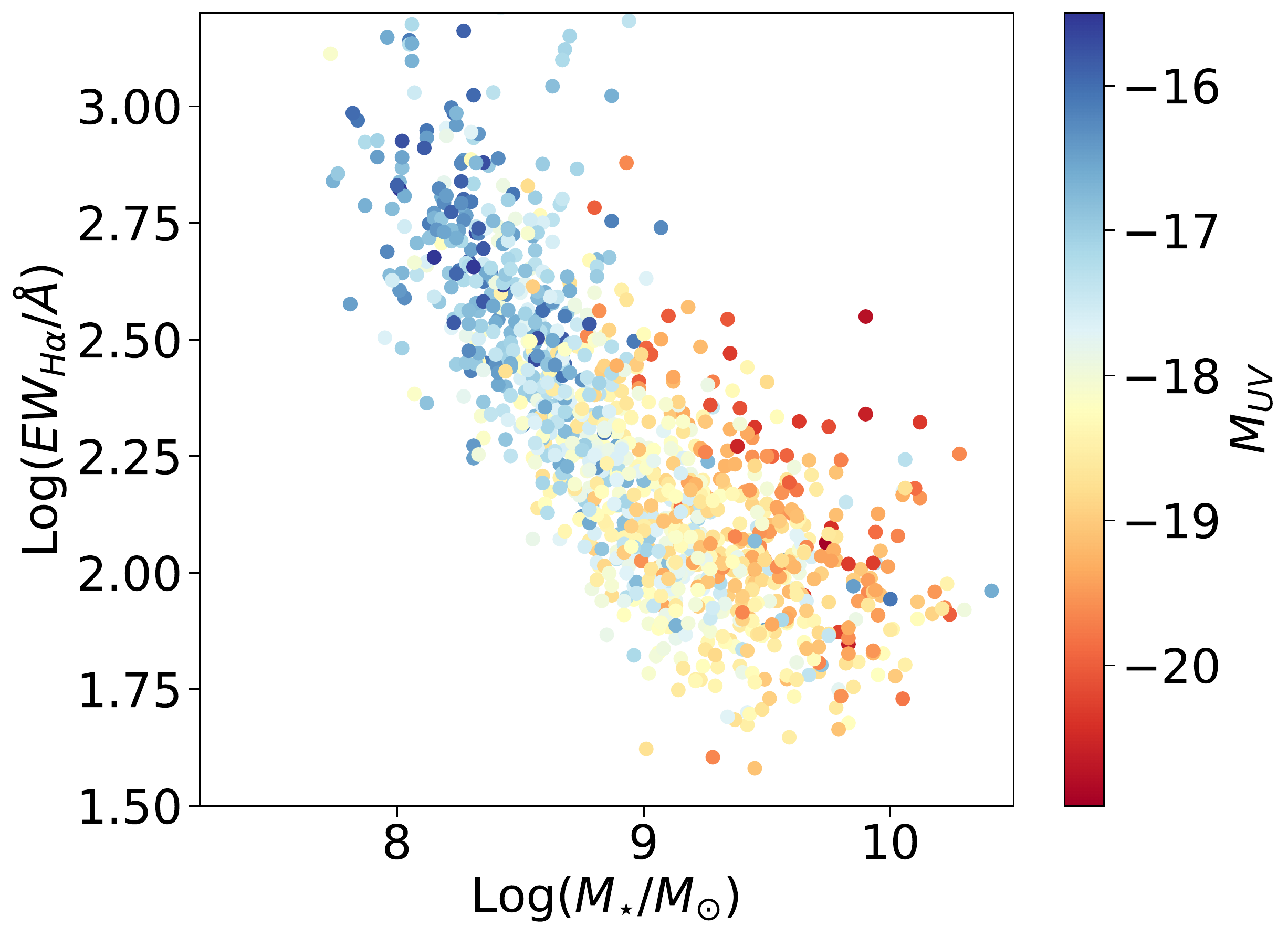}
    \caption{The rest-frame \ha\ equivalent width as a function of the stellar mass. The color code translates the absolute UV magnitude of the sample.}
    \label{fig:ewha_mass}
\end{figure}

\begin{figure}
        \includegraphics[width=8.5cm]{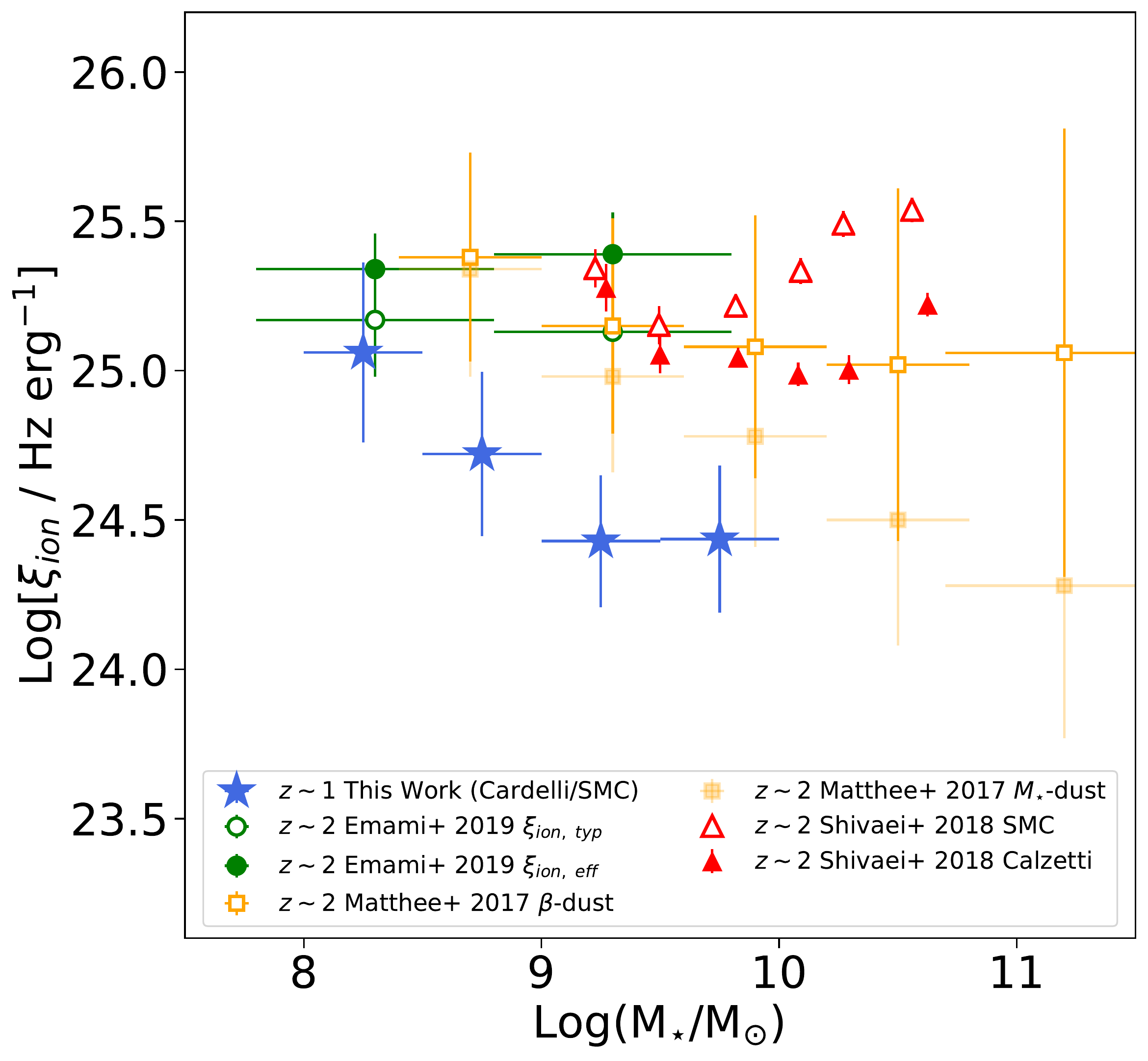}
    \caption{\xiion as a function of the stellar mass. Our \xiion\ measurements are represented with filled blue stars. These \xiion\ estimates are derived by correcting \ha\ line flux for nebular attenuation using the Cardelli curve and UV continuum luminosity for stellar extinction using SMC curve. Green filled (open) circles show the results of \citet{emami20} using the effective (standard) stacking method. The red filled (open) triangles show \xiion\ estimates by \citet{shivaei18} corrected for dust using Calzetti(SMC) curves. The results of \citet{matthee17} are shown with open, light-filled, and filled squares with dust corrections based on the $\beta$ slope, the mass-to-dust relation, and E(B-V) from SED fitting, respectively. More details on these literature results are given in Section \ref{sec:results}}
    \label{fig:xi_mass}
\end{figure}

As we have seen, \xiion\ appears to correlate well with the \ha\ equivalent width. At the same time, \ewha\ is anti-correlated with the stellar mass (cf. Figure \ref{fig:ewha_mass}). Similar results have been observed in galaxy samples at different redshifts \citep[e.g.][]{reddy18b,faisst19}. It is therefore expected that \xiion\ will correlate with the stellar mass. However, several studies have found that \xiion\ is generally independent of the stellar mass. In particular, over the same mass range as our sample, \citet{emami20} find no evolution of \xiion\ between their two stacks binned in mass. At higher masses, \citet{shivaei18} find a small increase of 0.23 dex towards lower masses in the case of a \citet{calzetti00} attenuation curve for the stellar continuum emission. 
Figure \ref{fig:xi_mass} shows how \xiion\ is related to the stellar mass in our sample, together with literature results at higher redshifts. We see a higher \xiion\ at lower mass. Because of the redshift-evolution we have seen before, the general trend is slightly offset from the $z \sim 2$ results. The observed trend is unlikely the result of mass completeness (cf. Section \ref{sec:sample}). However, while the sample is generally limited by the equivalent width selection of EW$=80$ \AA, it starts to be purely \ha-flux limited below $10^{9}$\msol (cf. Fig. \ref{fig:comp}). Such incompleteness could favor high \xiion\ values in lower-mass galaxies. Yet, we do not find high \xiion\ values in galaxies with stellar masses higher than $10^{9}$\msol, which is not affected by incompleteness. Deeper spectroscopic surveys of low-mass galaxies are required in order to confirm this trend. 

The largest uncertainty remains the dust correction, as small variation in the dust content or the attenuation law has a large impact on the derived \xiion, mainly because of the correction in the UV. We also over-plot the results of \citet{matthee17} with different dust correction methods, all of which show a trend of increase of \xiion\ towards lower masses, and highlight the dust-correction uncertainties. Note that, the $\beta$-based dust correction of \citet{matthee17} assumes $\beta_{0}=-2.21$ \citep{meurer99}, which contributes to the observed offset relative to our results.  

\begin{figure}
        \includegraphics[width=8.5cm]{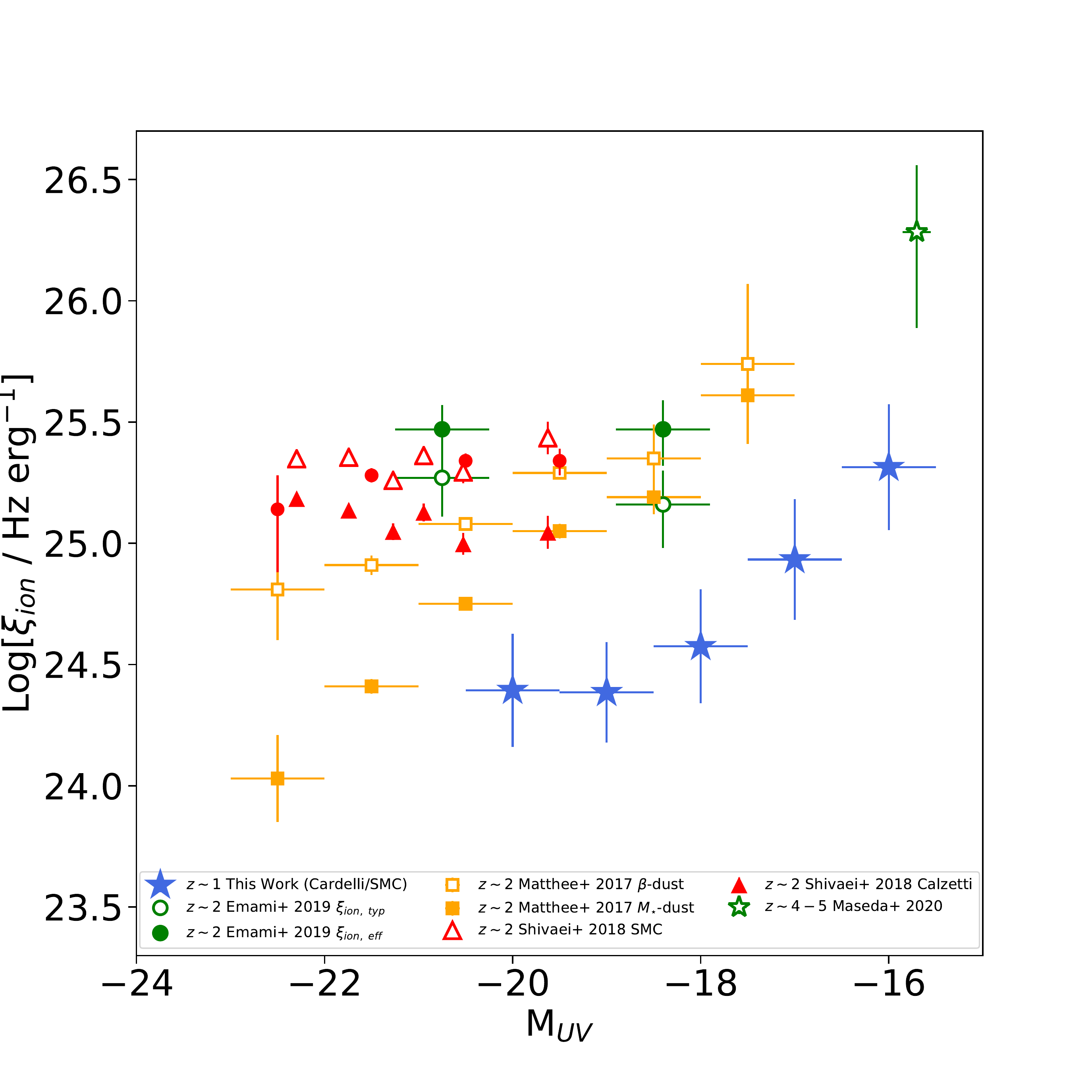}
    \caption{\xiion\ as a function of the absolute UV magnitude. Our \xiion\ estimates and literature results are represented with the same color code as in Figure \ref{fig:xi_mass}. In addition, the results of \citet{bouwens16} at $3.8 <z < 5.0$ are represented with red circles. Their measurements are inferred from IRAC-based \ha\ fluxes and corrected for dust attenuation using the $\beta$ slope and the SMC extinction curve. We also plot the result of \citet{maseda20} who similarly derived \xiion\ for a sample of Ly$\alpha$ emitters at $4<z<5$ using IRAC excess to estimate the \ha\ flux. No dust correction was applied.}
    \label{fig:xi_muv}
\end{figure}

We also investigate the relation between \xiion\ and the absolute UV magnitude, for which most studies do no report a clear correlation. As shown in Figure \ref{fig:xi_muv}, \citet{shivaei18} and \citet{emami20} report no evolution with $M_{UV}$, while \citet{matthee17} show an increase towards fainter magnitudes. Similarly to our previous result regarding the stellar mass, faint galaxies show significantly higher ionizing efficiency. At even fainter magnitudes than our sample, a stacking analysis of a sample of LAEs at $z=4-5$ by \citet{maseda20} shows \xiion\ falling along this trend. 

\section{Implications for reionization}
\label{sec:reionization}

Early star-forming galaxies are currently the best candidates for driving cosmic reionization \citep{robertson15, bouwens15,atek15b}. Given the most recent observational constraints on the shape of the UV luminosity function at $z>6$, low-mass galaxies could provide the bulk of the ionization budget \citep{livermore17,atek18, ishigaki18, finkelstein19}. However, precise knowledge of the ionizing properties of early galaxies and the escape fraction of ionizing radiation is still required to infer their exact contribution to reionization. Since such investigations are still challenging at $z >6$, many studies rely on lower-redshift galaxies, with the underlying assumption that they have the same properties as early galaxies. Therefore, it is important to focus on low-mass galaxies that are more representative of the galaxy population at the epoch of reionization.

It is clear that many physical properties and the SFH of low-mass galaxies are different from what is seen in their massive counterpart. The prevalence of intense emission lines, for instance, is much higher in low-mass galaxies, and increases with redshift. Star formation proceeds in successive bursts in low-mass galaxies, rather than a steady and constant star formation rate. Figure \ref{fig:sfrs_m} shows how the relative increase in the recent SFR, as traced by \ha\ relative to the averaged past SFR, as traced by the UV, indicate a recent burst of star formation in low-mass galaxies. While these observations cannot reveal successive SF bursts, many hydrodynamical simulations show that bursty star formation indeed happens mostly in low-mass galaxies, and can reproduce the observed variations of \ha/UV luminosity ratios. \citep{shen14,dominguez15,sparre17,emami19}. Confronting numerical simulations to the present observations is out of the scope of this paper and will be investigated in future work.
Overall, understanding the burstiness parameters in these galaxies is crucial for several reasons: 
\begin{itemize}
   \item While the canonical value of Log(\xiion) $=25.2$, based on a constant star formation, might be adequate for massive galaxies, such assumptions do not hold in low-mass galaxies. Hydrodynamical simulations have shown that bursty star formation history leads to large variations in \xiion\ compared to constant SFH \citep[e.g.][]{dominguez15}, which means the observed LyC photon production is not necessarily representative of the average value in these galaxies. For instance, these galaxies may remain undetected in LyC following a burst of star formation, although their average LyC production is high, and even if they have a high escape fraction. This is because it takes about 5 Myr for LyC emission to decrease by an order of magnitude after the burst, whereas the escape fraction will be significant only after 10 Myr, which corresponds to the supernovae timescale. However, accounting for binary stars in stellar population models \citep[e.g.][]{stanway18} lead to a significant increase in the number of ionizing photons produced over a larger range of masses, which result in a significant ionizing emission beyond 10 Myr \citep[e.g.][]{chisolm19}.   
    
    \item  Bursty star formation also means that star formation feedback, which is responsible for disrupting the neutral ISM and clearing sight lines, will have significant impact on the LyC escape fraction. Hydrodynamical simulations also find large time variations of the escape fraction in low-mass galaxies, driven by supernovae explosions of massive stars \citep{kimm14, trebitsch17}. We note that in estimating \xiion, we assumed \fesc $=0$. This is only a convention to ensure a fair comparison with literature results and the redshift-evolution of \xiion. The most recent observing campaigns show a wide range of values for \fesc\ at different redshifts \citep{izotov16,izotov18,shapley16,vanzella18,flury22}. While empirically constraining the evolution of \fesc\ remains challenging, hydrodynamical simulations find correlations with galaxy properties. For example, \fesc\ increases with increasing star formation efficiency (or sSFR) and decreasing metallicity \citep[e.g.][]{kimm19}.
\end{itemize}
   
These two quantities are pivotal in estimating the total ionizing power of galaxies. In addition to their dependency with galaxy physical parameters, these quantities might also be intertwined, as a higher ionizing photon production is likely followed by an increase in the escape fraction. Both their average values seem to increase with redshift, indicating a higher ionizing power in early galaxies than their low$-z$ analogs.

\section{Conclusions}
\label{sec:conclusions}

In this work, we used a combination of {\em HST} grism spectroscopy from the  3DHST survey and UV imaging from the HDUV survey for a large sample of 1167 galaxies across a redshift range of $0.7<z< 1.5$. This represents a unique sample with its combination of direct measurements of \ha\ down to an emission line flux limit of $1.5 \times 10^{-17}$ \ergscm\ and deep UV imaging down to $M \sim 28$ mag. This combination enables us to probe star formation and UV output down to low stellar masses between $\sim 10^{8}$ and $10^{10}$ \msol. We have explored the star formation histories of these low-mass galaxies and estimated their ionizing efficiency \xiion. In an attempt to extrapolate these findings to higher-redshift galaxies, we also explored how these parameters evolve with the physical properties of galaxies and with redshift. 

We used the \ha\ emission to infer the ionizing photon production rate, which in turn was combined with the UV luminosity at 1500 \AA\ to compute the ionizing efficiency \xiion. For removing the attenuation due to dust, we used the Cardelli curve to correct the \ha\ flux and the SMC curve to correct the UV luminosity. With this combination, we obtained a  median ionizing efficiency of Log(\xiion/erg$^{-1}$ Hz)$=24.80 \pm 0.26$. While adopting different curves, the nebular attenuation has little impact on \xiion\, the UV attenuation is a source of significant uncertainty. For example, adopting \citet{calzetti00} law for the UV attenuation reduces the above median \xiion\ by 0.54 dex. Overall, our determination is significantly lower than literature results at higher redshifts. We find a clear evolution with lookback time, ranging from Log(\xiion)$=24.70$ at $t_{L} \sim 8.3$ Gyr to Log(\xiion)$=25.54$ at $t_{L} \sim  12.7$ Gyr. This evolution can be explained by higher star formation efficiency and higher emission line EWs on average at higher redshifts.

Investigating the relation between \xiion\ and galaxy properties:
\begin{itemize}
    \item We observe a strong correlation with the \ha\ equivalent width, consistent with the previous findings at different redshifts \citep{tang19,matthee17, reddy18b, emami19}.
    
    \item We found that \xiion\ is anti-correlated with stellar mass. This result should not be surprising since we also observe a strong anti-correlation between \ewha\ and stellar mass. However, incompleteness at the low-mass end likely affects the slope of the correlation. Previous studies found that \xiion\ is mostly independent of the stellar mass \citep{shivaei18,emami19}.
    
    \item Similarly, there is an evolution with absolute UV magnitude where fainter galaxies show higher \xiion\ values. Studies by \citet{matthee17} and \citet{maseda20} show results consistent with this trend.

\end{itemize}

When considering the SFR-\mstar\ relation, we find that a significant fraction of our sample lies above the main sequence, established at a similar redshift \citep{whitaker14, schreiber15}. Only galaxies around \ewha\ $\sim 100$ \AA\ follow the MS, while higher \ewha\ galaxies depart from the MS, presumably because of an elevated instantaneous SFR following a recent burst. This is particularly true for lower-mass galaxies, below 10$^{9}$ \msol, where high EWs are more common and contribute to a larger dispersion in the MS relation. 

This result is confirmed when comparing the \ha\ and UV continuum SFR indicators as a function of mass. We observe high \sfrha/\sfruv\ ratios in galaxies with $10^{9}$ \msol. Unsurprisingly, these galaxies also show the highest \ewha. Given the selection bias towards strong \ha\ emitters in low-mass galaxies, these results can also be characterized as a higher scatter in the \sfrha/\sfruv\ distribution in lower-mass galaxies, as it has been observed in previous studies \citep{weisz12,guo16,sparre17,emami19}. \citet{faisst19} results show that 50\% of galaxies show a significant excess (more than a factor of 2) in SFR derived from \ha\ with respect to the equilibrium. These galaxies also show the highest \ha\ equivalent widths. In general, there is a large scatter around the equilibrium value of \sfrha/\sfruv.  

On the other hand, \citet{broussard19} compare theoretical predictions from semi-analytical models and hydrodynamical simulations to 3D-HST observations to infer the SF burstiness. They define the burstiness parameter $\eta$ as the logarithm of the ratio between \sfrha\ and \sfruv. They find a positive correlation between the average of $\eta$ and stellar mass, and the scatter appears to be independent of the stellar mass. They note however that this correlation could be due to systematics in 3D-HST data that are not implemented in mock galaxies or to the dust prescriptions used. Our findings are in disagreement with those results, since we find that (i) the \sfrha/\sfruv\ increases towards lower masses, even when using the same dust prescription to correct the \ha\ and the UV emissions (Bottom-right panel of Figure \ref{fig:app2}); or (ii) given the selection effects (see Section \ref{sec:sfrs_m}) the scatter of \sfrha/\sfruv\ ratio increases towards lower masses. This disagreement could be due to differences in both the sample selection and the methodology. Firstly, it is unclear whether the population of low-mass galaxies (fainter than H=24 AB mag) used in \citet{broussard19} has gone through additional checks including visual inspection, as it has been done in the present study. Indeed, this sub-sample of galaxies in the 3D-HST catalog is prone to uncertain measurements and spurious spectral identifications. Secondly, the SED fitting procedure and parameters used to derive their stellar population properties are different from the ones used in our study.

Some theoretical studies also show that burstiness is common in high-redshift galaxies across a wide mass range, whereas others show that only low-mass galaxies remain bursty all the way down to $z=0$ \citep{muratov15, velasquez20}). Our results suggest that burstiness parameters such as the duration of the burst or the duty cycle can differ from theoretical predictions. The prevalence of galaxies with enhanced \ha\ relative to UV emission requires star formation variation on short time scales (less than 10 Myr) and/or short idle time between burst episodes on a rapid duty cycle. Direct implications on models of galaxy formation include modifications in the balance between the infall of gas, on one hand, and the ejection of gas or photoionization by the UV background, on the other hand. 

Finally, a realistic model of reionization requires the convolution of the galaxy UV LF with physically-motivated ionization properties of galaxies. Recent attempts based on current observations lead to a variety of conclusions regarding the nature of galaxies contributing the most to reionization \citep[e.g.][]{finkelstein19, naidu20}. A better understanding of the burstiness of galaxies and an accurate assessment of the volumetric escape fraction of galaxies (e.g. as a function of stellar mass) through deep LyC and rest-frame optical observations are needed to improve on this front. 

Future NIR spectroscopic observations with the upcoming {\em JWST} will help obtain accurate measurements of the SFR and dust attenuation through the Balmer decrement for a large sample of galaxies across a wide range of redshifts to investigate in details the SF burstiness in galaxies down to even lower masses.

\section*{Acknowledgements}

 We thank Irene Shivaei, Najmeh Emami, and Jorryt Matthee for sharing their tabulated data. HA thanks Dan Stark and Brian Siana for useful discussions. This work is supported by CNES. It is based on observations obtained with the NASA/ESA Hubble Space Telescope, obtained from  the Mikulski Archive for Space Telescopes at the Space Telescope Science Institute, which is operated by the Association of Universities for Research in Astronomy, Inc., under NASA contract NAS 5-26555. These observations are associated with programs 12177, 11600, 12534, and 13872. PAO acknowledges support from the Swiss National Science Foundation through the SNSF Professorship grant 190079 `Galaxy Build-up at Cosmic Dawn'. The Cosmic Dawn Center (DAWN) is funded by the Danish National Research Foundation under grant No.\ 140.
 
 \section*{Data Availability}
 The data underlying this article are publicly available in the MAST archive at \url{https://archive.stsci.edu/prepds/3d-hst/} and \url{https://archive.stsci.edu/prepds/hduv/}.
 


\bibliographystyle{mnras}
\bibliography{references}



\appendix

\section{Luminosity completeness}
\label{app:a1}
\begin{figure}
    \centering
    \includegraphics[width=8cm]{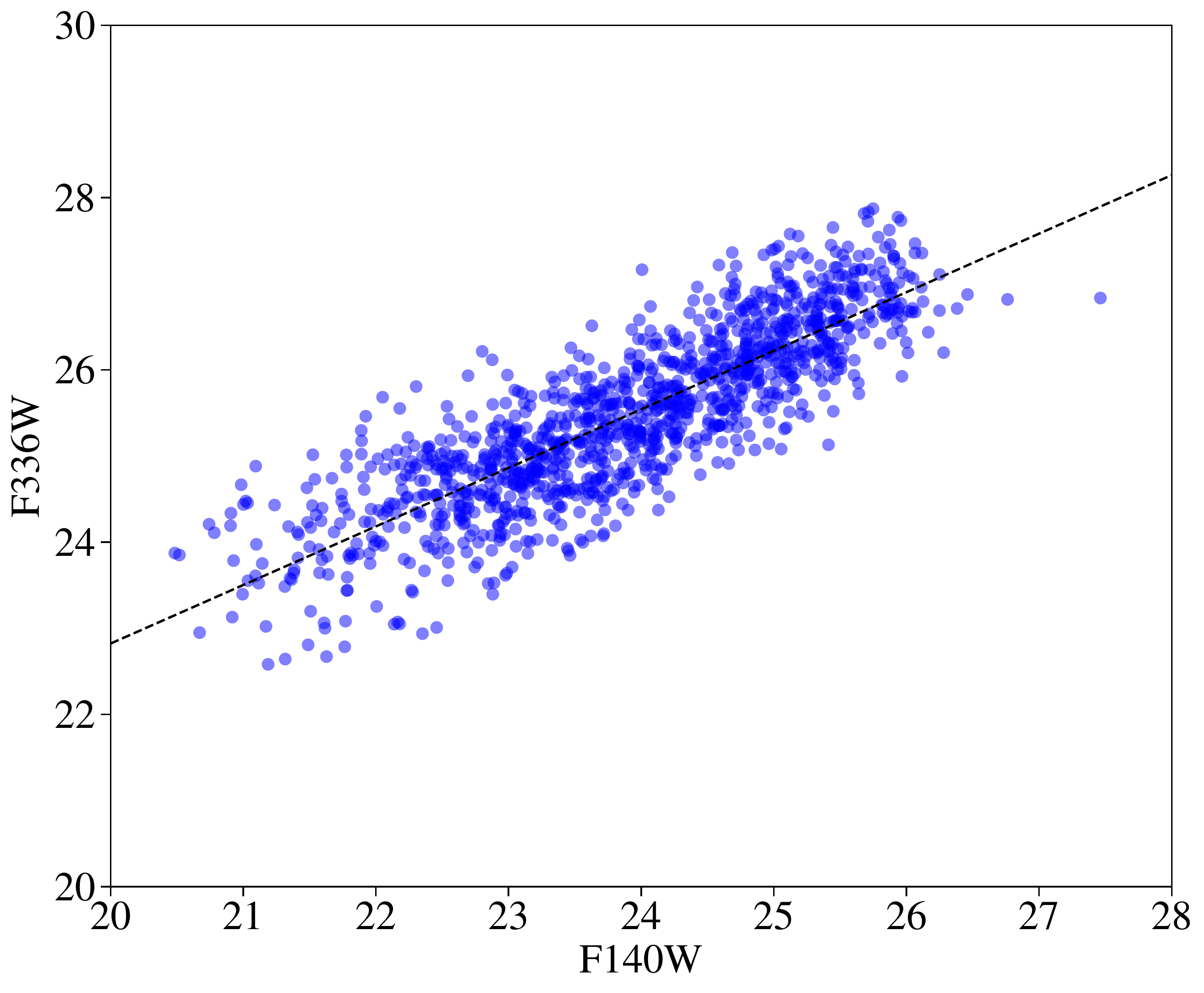}
    \caption{The UV magnitude in the F336W filter versus the NIR magnitude in the F140W filter for the \xiion\ sample. The dashed line shows the best-fit relation}
    \label{fig:app1}
\end{figure}

In section \ref{sec:xi}, in order to estimate the completeness of our sample in terms of \xiion\ measurements, we combined the \ha\ flux limit and the \ewha. We simulated a sample of \ewha\ following the observed distribution, applying the same selection cut at 80\AA. A combination with the \ha\ flux limit gives a distribution in F140W magnitudes (rest-frame optical continuum). For the second term \luv, we use the observed correlation between the UV (F336W) and NIR (F140W) magnitudes and its associated dispersion. Figure \ref{fig:app1} shows this relation, with a correlation of $UV_{mag} = (0.68 \pm 0.012) \times H_{mag}  + (9.22 \pm 0.29) $

\section{Dust corrections}
\label{app:a2}

To explore how the choice of dust laws affects the burstiness results, we computed the \sfrha/\sfruv\ ratio as a function of stellar mass for four different combinations of nebular/stellar corrections. The results are shown in Figure \ref{app:a2}. We first adopt the Calzetti attenuation law for both the \sfrha\ and \sfruv\ (Top-left panel). As we have explained in Section \ref{sec:dust} (see also Table \ref{tab:dust}), changing the dust law leads to significant difference in the correction factors mainly in the UV domain. Therefore, the adoption of the Calzetti law for the UV makes the \sfruv\ significantly larger, especially for the more massive galaxies, which makes the \sfrha/\sfruv\ much lower. The net result is a stronger dependence of the burstiness with stellar mass. We also investigate the following combinations: SMC/SMC and the Calzetti/Calzetti, but this time deriving the nebular attenuation from the stellar one, assuming a ratio of  \ebvs/\ebvg=0.44. The fiducial combination Cardelli/SMC is also shown. These two combinations have little impact on the general trend of the SFR ratio, apart from slightly elevating the normalization of the bulk of the sample.

    \begin{figure*}
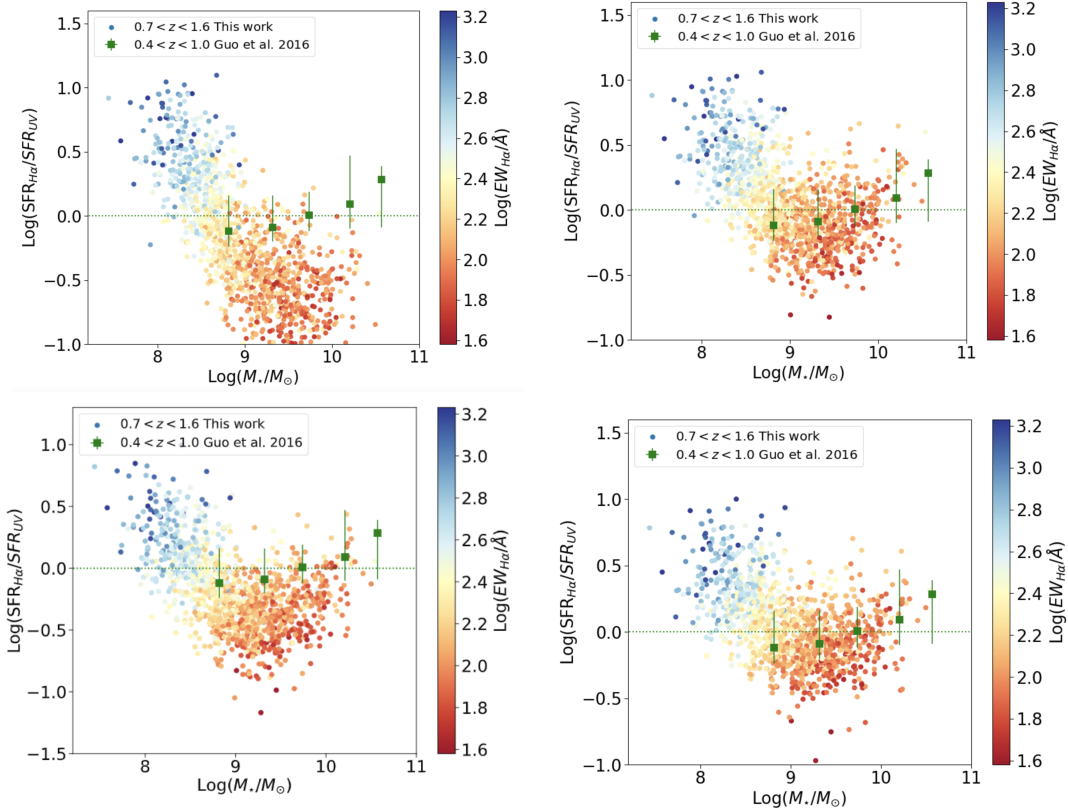

            \includegraphics[width=0.8\columnwidth]{sfrs_m_calz_calz.pdf}
            \hspace{0.3cm}
            \includegraphics[width=0.8\columnwidth]{sfrs_m_smc_smc.pdf}
            \hfill
            \includegraphics[width=0.8\columnwidth]{sfrs_m_card_smc.pdf}
             \hspace{0.3cm}
            \includegraphics[width=0.82\columnwidth]{sfrs_m_calzfact_calz.pdf}
        \caption{Effects of dust laws on the variations of the \sfrha/\sfruv\ as a function of stellar mass. Four combinations for nebular/stellar dust are tested here. {\em Top-left}: Calzetti/Calzetti, {\em top-right:} SMC/SMC, {\em Bottom-left:} Cardelli/SMC, {\em Bottom-right:} Calzetti/Calzetti assuming \ebvs=0.44\ebvg. The color code is the same as in Figure \ref{fig:sfrs_m}.
        } 
        \label{fig:app2}
    \end{figure*}

\section{Intrinsic UV slope}
\label{app:a3}

\begin{figure*}
    \centering
    \includegraphics[width=0.8\columnwidth]{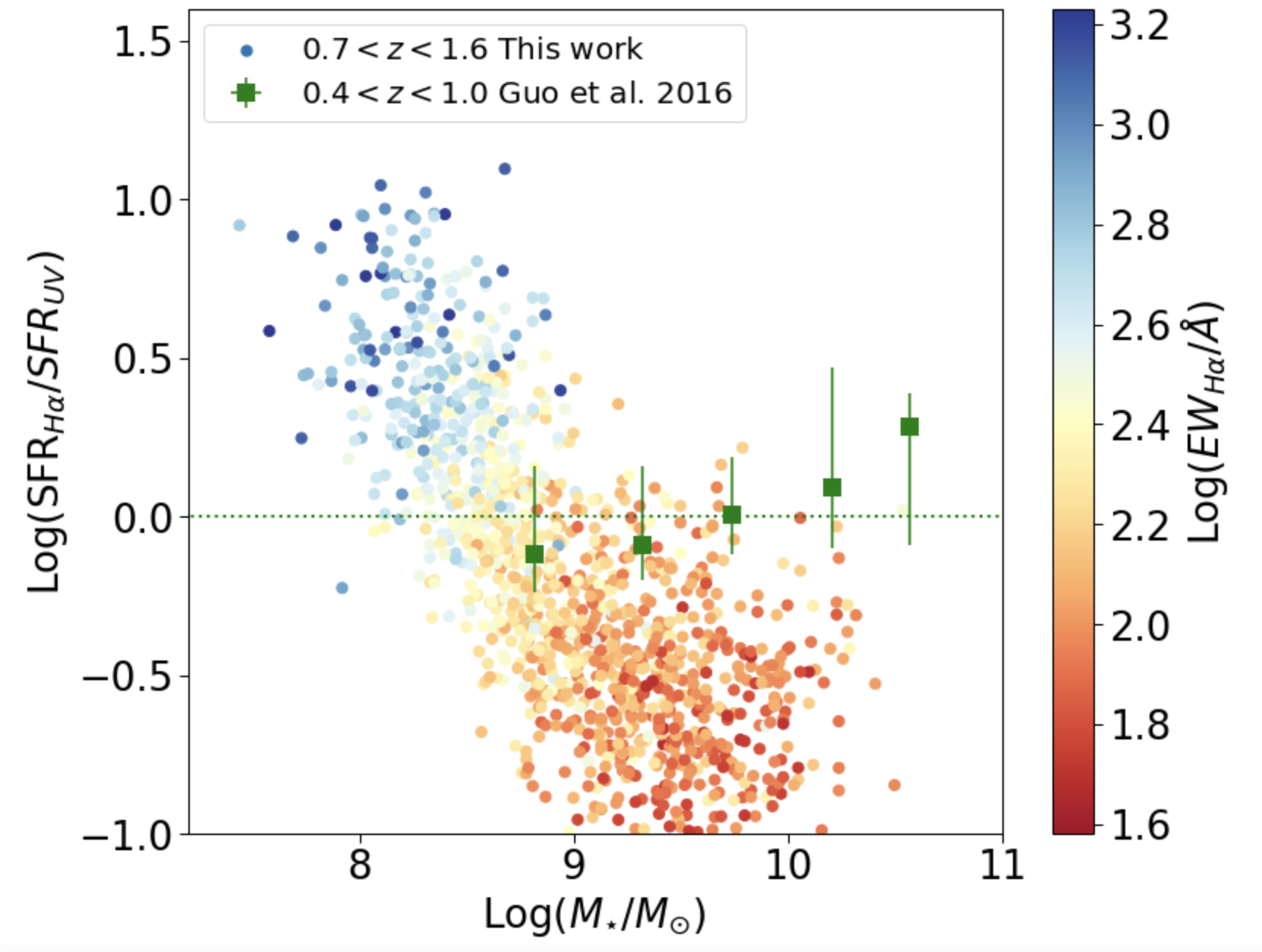}
    \includegraphics[width=0.8\columnwidth]{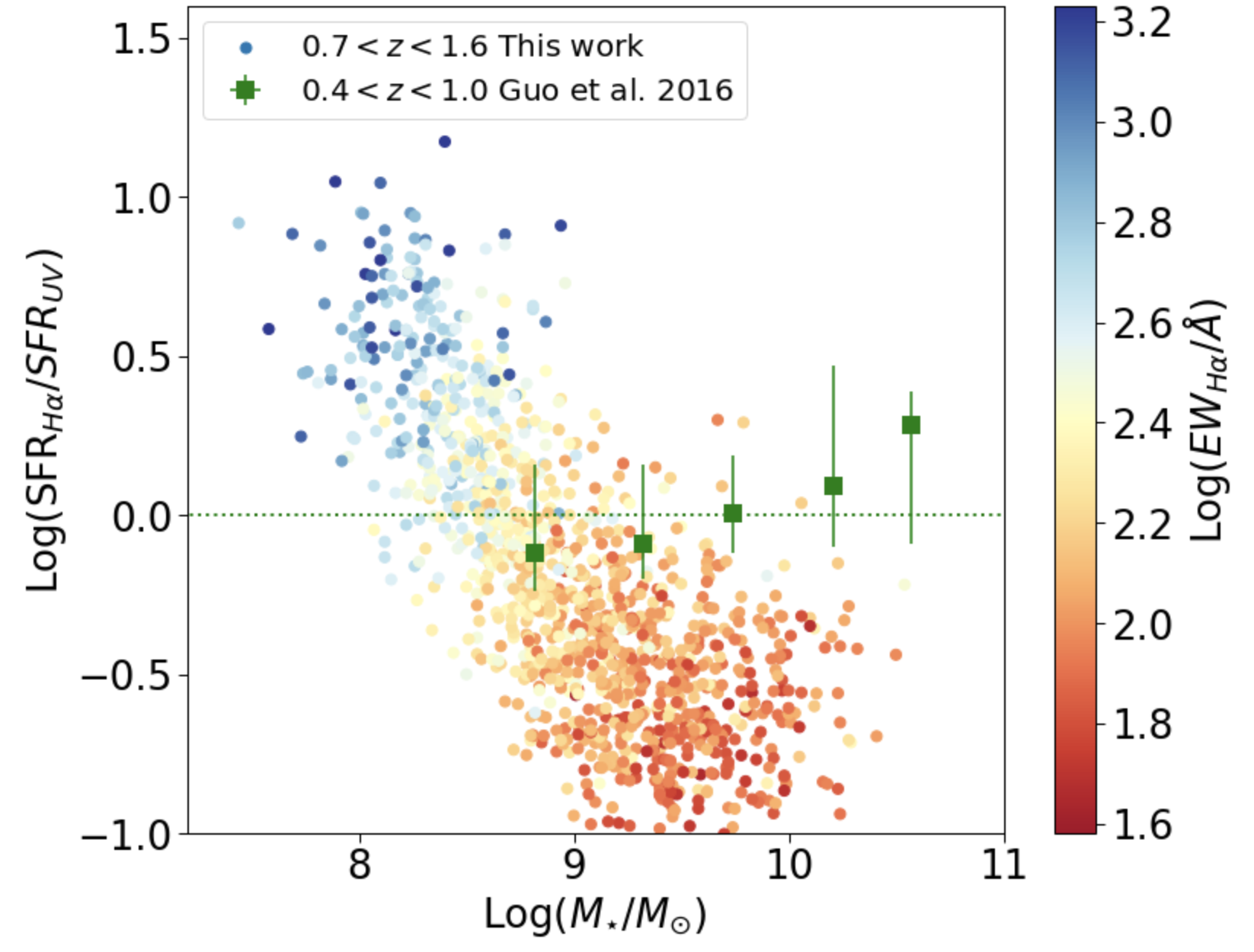}
    \caption{The impact of the intrinsic UV slope $\beta_{int}$ on the correlation between the \sfrha/\sfruv\ and the stellar mass. We compare, in the left panel, the value adopted in this study $\beta_{0}=-2.62$ \citep{reddy18} and, in the right panel, the canonical value of $\beta_{0}=-2.21$ \citep{meurer99}. We used the \citet{calzetti00} attenuation curve for both the stellar and nebular dust correction (cf. fig. \ref{fig:app2}).}
    \label{fig:app3a}
\end{figure*}

In computing the dust extinction of the UV stellar continuum, we have measured and compared the observed UV slope $\beta$ with the intrinsic slope $\beta_{0}$. We adopted a value of $\beta_{0}=-2.62$ derived by \citet{reddy18}. This value is derived for a sample of $z\sim 2$ galaxies, assuming a stellar population model with an age of 100 Myr, $Z=0.14Z_{\odot}$ metallicity and a constant star formation history. To investigate the importance of this parameter choice, we also used the canonical value of $\beta_{0}=-2.23$ \citep{meurer99} for comparison, which assumes a constant star formation history and a solar metallicity. In (Figure \ref{fig:app3a}), we show the effect on the \sfrha/\sfruv\ ratio as a function of stellar mass. We observe a slight elevation of the global values, with no noticeable impact on the overall trend. In \citet{meurer99}, it is also shown that variations of the stellar populations properties, including instantaneous bursts of star formation, IMF and metallicity lead to variations between $\beta_{0}=-2.0$ and $\beta_{0}=-2.6$.      

\begin{figure*}
    \centering
    \includegraphics[width=0.8\columnwidth]{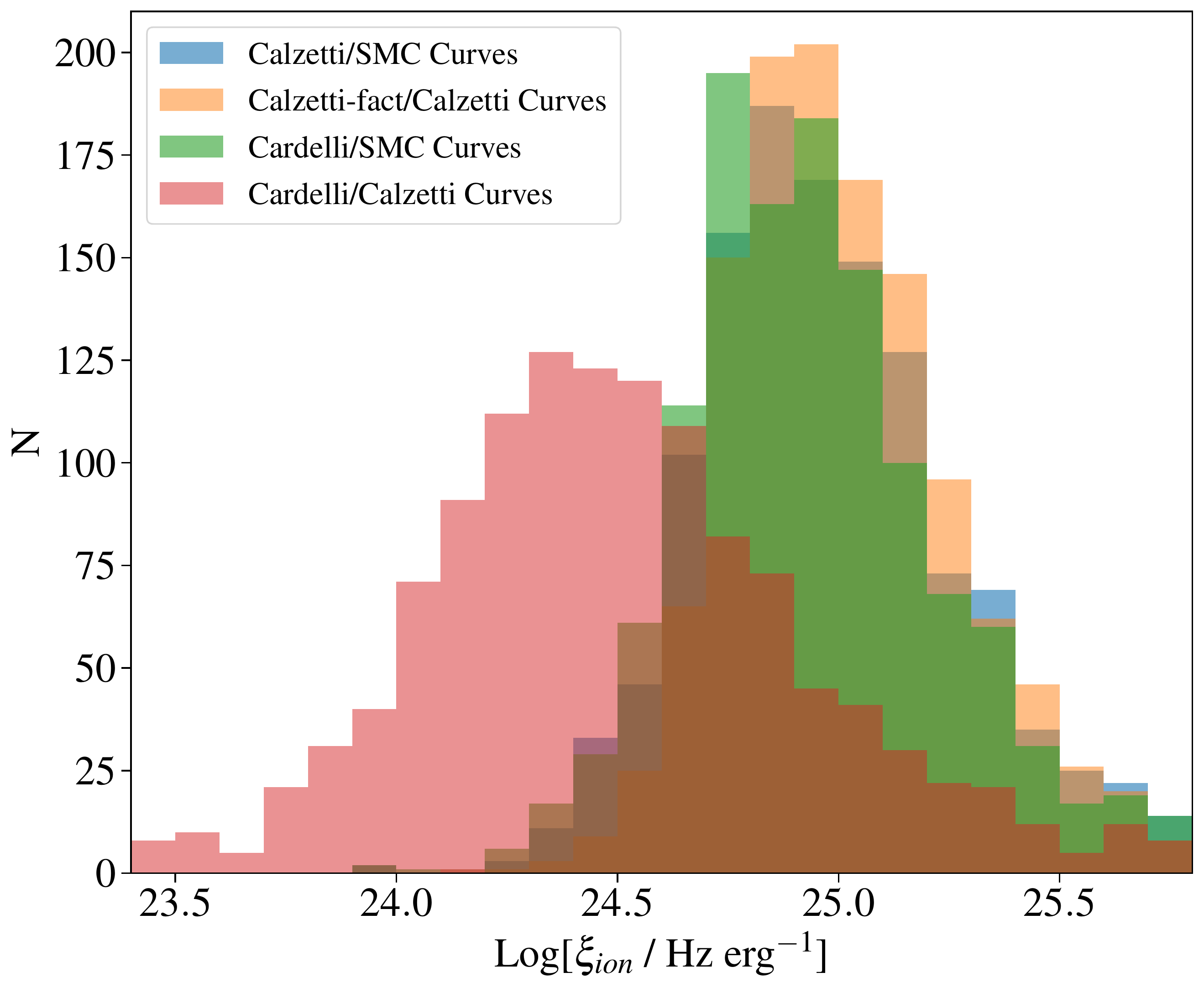}
    \includegraphics[width=0.8\columnwidth]{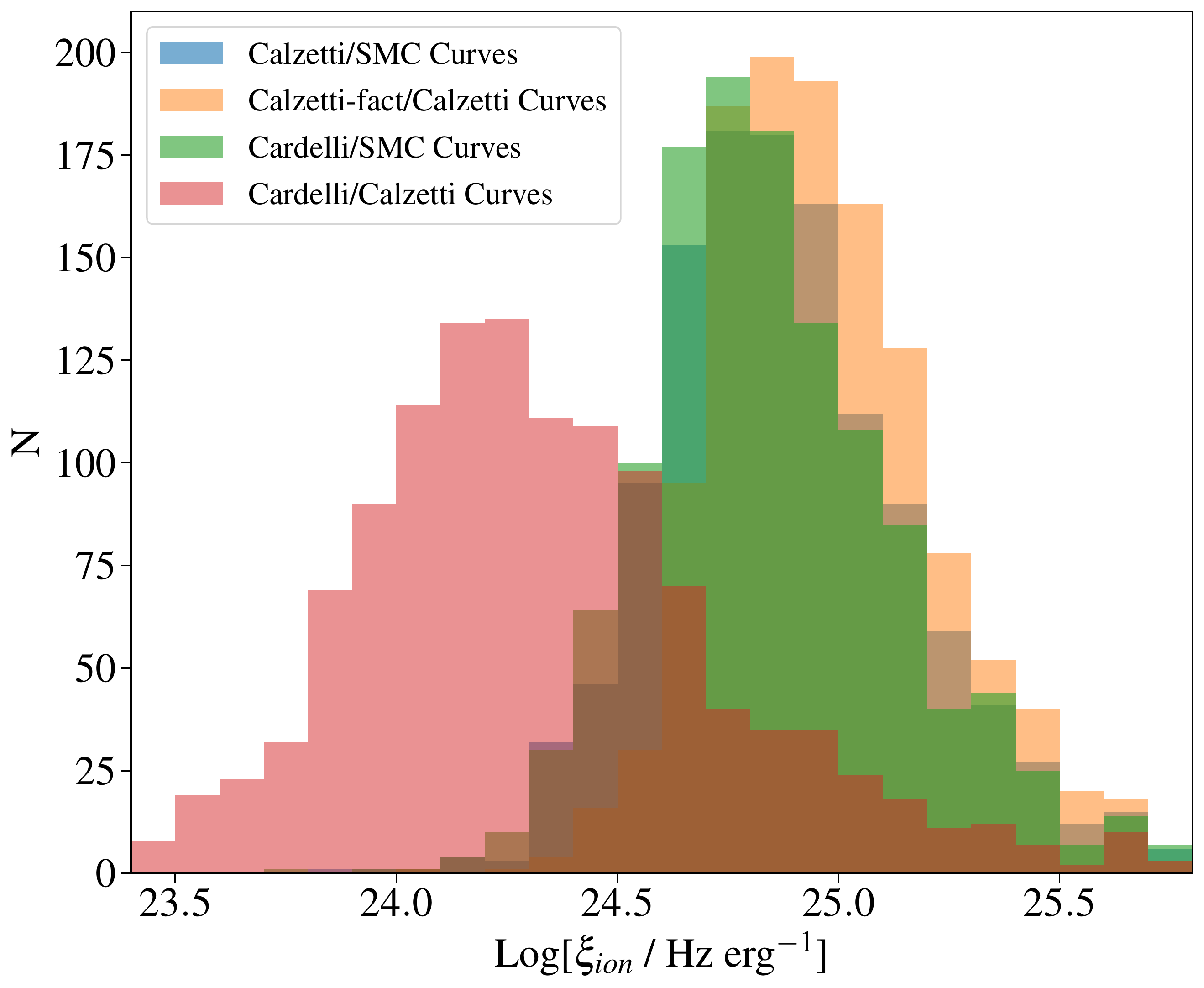}
    \caption{The impact of the intrinsic UV slope $\beta_{int}$ on \xiion\ distribution. The left panel shows the same distributions shown in figure \ref{fig:histo_xi} but adopting $\beta_{int} = -2.23$ \citep{meurer99} instead of $\beta_{int} = -2.62$ \citep{reddy18}. The right panel shows the result of using a dispersion in $\beta_{int}$ in the range defined by these two values.}
    \label{fig:app3b}
\end{figure*}

Similarly, we also show in Figure \ref{fig:app3b} the impact of adopting a different intrinsic UV slope on the \xiion\ distribution. First, adopting $\beta_{int} =-2.23$ results in a median value of Log(\xiion/erg$^{-1}$ Hz)$=25.12 \pm 0.26$, adopting the fiducial dust correction combination Cardelli/SMC. For reference, we found  Log(\xiion/erg$^{-1}$ Hz)$=24.80 \pm 0.26$ when using $\beta_{int} =-2.62$. We also included a dispersion in the intrinsic UV slope by uniformly sampling the range $-2.62<\beta < -2.23$, resulting in the right panel of Figure \ref{fig:app3b}. We find a median value of Log(\xiion/erg$^{-1}$ Hz)$=24.89 \pm 0.27$. A slight increase of 0.05 dex in the dispersion of the \xiion\ distribution  is notably observed for the Cardelli/Calzetti durst correction.

\bsp	
\label{lastpage}
\end{document}